\documentclass[%
reprint,
superscriptaddress,
nofootinbib,
nobibnotes,
 amsmath,amssymb,
 aps,
 prd,
 longbibliography,
 twocolumn,
]{revtex4-1}

\usepackage{siunitx}

\usepackage{graphicx}
\usepackage{dcolumn}
\usepackage{bm}
\usepackage{xcolor}
\usepackage[hidelinks]{hyperref}
\hypersetup{backref=true,       
    pagebackref=true,               
    hyperindex=true,                
    colorlinks=true,                
    breaklinks=true,                
    urlcolor= black,                
    linkcolor= red,                
    bookmarks=true,                 
    bookmarksopen=false,
    filecolor=black,
    citecolor=blue,
    linkbordercolor=blue
}
\usepackage{mathtools}
\usepackage{amsmath}
\usepackage{breqn}
\usepackage{comment}
\usepackage{multirow}
\usepackage{caption}
\usepackage{subcaption}
\usepackage{float}
\usepackage{makecell}
\usepackage{overpic}
\usepackage{breqn}
\usepackage[labelformat=simple]{subcaption}
\usepackage{appendix}

\usepackage{soul}
\usepackage{xspace}
\usepackage{booktabs}

\usepackage{layouts}
\usepackage{fancyhdr}

\fancypagestyle{titlepage}{\thispagestyle{fancy}}

\newlength{\figwidth}
\newlength{\fighalfwidth}
\newcommand{\OverallModel}{overall simulation model\xspace}

\DeclareSIUnit \mm {\milli\meter}
\DeclareSIUnit \cm {\centi\meter}
\DeclareSIUnit \us {\micro\second}
\DeclareSIUnit \ms {\milli\second}
\DeclareSIUnit \pA {\pico\ampere}
\DeclareSIUnit \pC {\pico\coulomb}
\DeclareSIUnit \fC {\femto\coulomb}
\DeclareSIUnit \fF {\femto\farrad}
\DeclareSIUnit \pF {\pico\farrad}
\DeclareSIUnit \mV {\milli\volt}
\DeclareSIUnit \kV {\kilo\volt}
\DeclareSIUnit \V {\volt}
\DeclareSIUnit \GOhm {\giga\ohm}
\DeclareSIUnit \MOhm {\mega\ohm}
\DeclareSIUnit \ton {\tonne}
\DeclareSIUnit \kton {\kilo\tonne}
\DeclareSIUnit \kt {\kilo\tonne}
\DeclareSIUnit \Mt {\mega\tonne}
\DeclareSIUnit \eV {\electronvolt}
\DeclareSIUnit \keV {\kilo\electronvolt}
\DeclareSIUnit \MeV {\mega\electronvolt}
\DeclareSIUnit \GeV {\giga\electronvolt}
\DeclareSIUnit \km {\kilo\meter}
\DeclareSIUnit \kW {\kilo\watt}
\DeclareSIUnit \MW {\mega\watt}
\DeclareSIUnit \MHz {\mega\hertz}
\DeclareSIUnit \kHz {\kilo\hertz}
\DeclareSIUnit \mrad {\milli\radian}
\DeclareSIUnit \year {year}
\DeclareSIUnit \POT {POT}
\DeclareSIUnit \sig {$\sigma$}
\DeclareSIUnit\parsec{pc}
\DeclareSIUnit\lightyear{ly}
\DeclareSIUnit\foot{ft}
\DeclareSIUnit\ft{ft}

\newcommand{\uboone}{MicroBooNE\xspace}
\newcommand{\vlne}{RNN\xspace}

\newcommand{\sref}[1]{\hyperref[#1]{Sec.~\ref*{#1}}}
\newcommand{\Sref}[1]{\hyperref[#1]{Section~\ref*{#1}}}

\newcommand{\aref}[1]{\hyperref[#1]{Appendix~\ref*{#1}}}
\newcommand{\Aref}[1]{\hyperref[#1]{Appendix~\ref*{#1}}}

\newcommand{\fref}[1]{\hyperref[#1]{Fig.~\ref*{#1}}}
\newcommand{\Fref}[1]{\hyperref[#1]{Figure~\ref*{#1}}}

\begin{document}

\raggedbottom

\title{
    Improving neutrino energy estimation of charged-current interaction events  with recurrent neural networks in MicroBooNE
}

\newcommand{\ANL}{Argonne National Laboratory (ANL), Lemont, IL, 60439, USA}
\newcommand{\Bern}{Universit{\"a}t Bern, Bern CH-3012, Switzerland}
\newcommand{\BNL}{Brookhaven National Laboratory (BNL), Upton, NY, 11973, USA}
\newcommand{\UCSB}{University of California, Santa Barbara, CA, 93106, USA}
\newcommand{\Cambridge}{University of Cambridge, Cambridge CB3 0HE, United Kingdom}
\newcommand{\CIEMAT}{Centro de Investigaciones Energ\'{e}ticas, Medioambientales y Tecnol\'{o}gicas (CIEMAT), Madrid E-28040, Spain}
\newcommand{\Chicago}{University of Chicago, Chicago, IL, 60637, USA}
\newcommand{\Cincinnati}{University of Cincinnati, Cincinnati, OH, 45221, USA}
\newcommand{\CSU}{Colorado State University, Fort Collins, CO, 80523, USA}
\newcommand{\Columbia}{Columbia University, New York, NY, 10027, USA}
\newcommand{\Edinburgh}{University of Edinburgh, Edinburgh EH9 3FD, United Kingdom}
\newcommand{\FNAL}{Fermi National Accelerator Laboratory (FNAL), Batavia, IL 60510, USA}
\newcommand{\Granada}{Universidad de Granada, Granada E-18071, Spain}
\newcommand{\Harvard}{Harvard University, Cambridge, MA 02138, USA}
\newcommand{\IIT}{Illinois Institute of Technology (IIT), Chicago, IL 60616, USA}
\newcommand{\Indiana}{Indiana University, Bloomington, IN 47405, USA}
\newcommand{\KSU}{Kansas State University (KSU), Manhattan, KS, 66506, USA}
\newcommand{\Lancaster}{Lancaster University, Lancaster LA1 4YW, United Kingdom}
\newcommand{\LANL}{Los Alamos National Laboratory (LANL), Los Alamos, NM, 87545, USA}
\newcommand{\Louisiana}{Louisiana State University, Baton Rouge, LA, 70803, USA}
\newcommand{\Manchester}{The University of Manchester, Manchester M13 9PL, United Kingdom}
\newcommand{\MIT}{Massachusetts Institute of Technology (MIT), Cambridge, MA, 02139, USA}
\newcommand{\Michigan}{University of Michigan, Ann Arbor, MI, 48109, USA}
\newcommand{\MSU}{Michigan State University, East Lansing, MI 48824, USA}
\newcommand{\Minnesota}{University of Minnesota, Minneapolis, MN, 55455, USA}
\newcommand{\Nankai}{Nankai University, Nankai District, Tianjin 300071, China}
\newcommand{\NMSU}{New Mexico State University (NMSU), Las Cruces, NM, 88003, USA}
\newcommand{\Oxford}{University of Oxford, Oxford OX1 3RH, United Kingdom}
\newcommand{\Pitt}{University of Pittsburgh, Pittsburgh, PA, 15260, USA}
\newcommand{\Rutgers}{Rutgers University, Piscataway, NJ, 08854, USA}
\newcommand{\SLAC}{SLAC National Accelerator Laboratory, Menlo Park, CA, 94025, USA}
\newcommand{\SDSMT}{South Dakota School of Mines and Technology (SDSMT), Rapid City, SD, 57701, USA}
\newcommand{\Maine}{University of Southern Maine, Portland, ME, 04104, USA}
\newcommand{\Syracuse}{Syracuse University, Syracuse, NY, 13244, USA}
\newcommand{\TelAviv}{Tel Aviv University, Tel Aviv, Israel, 69978}
\newcommand{\Tennessee}{University of Tennessee, Knoxville, TN, 37996, USA}
\newcommand{\UTA}{University of Texas, Arlington, TX, 76019, USA}
\newcommand{\Tufts}{Tufts University, Medford, MA, 02155, USA}
\newcommand{\UCL}{University College London, London WC1E 6BT, United Kingdom}
\newcommand{\VTech}{Center for Neutrino Physics, Virginia Tech, Blacksburg, VA, 24061, USA}
\newcommand{\Warwick}{University of Warwick, Coventry CV4 7AL, United Kingdom}
\newcommand{\Yale}{Wright Laboratory, Department of Physics, Yale University, New Haven, CT, 06520, USA}

\affiliation{\ANL}
\affiliation{\Bern}
\affiliation{\BNL}
\affiliation{\UCSB}
\affiliation{\Cambridge}
\affiliation{\CIEMAT}
\affiliation{\Chicago}
\affiliation{\Cincinnati}
\affiliation{\CSU}
\affiliation{\Columbia}
\affiliation{\Edinburgh}
\affiliation{\FNAL}
\affiliation{\Granada}
\affiliation{\Harvard}
\affiliation{\IIT}
\affiliation{\Indiana}
\affiliation{\KSU}
\affiliation{\Lancaster}
\affiliation{\LANL}
\affiliation{\Louisiana}
\affiliation{\Manchester}
\affiliation{\MIT}
\affiliation{\Michigan}
\affiliation{\MSU}
\affiliation{\Minnesota}
\affiliation{\Nankai}
\affiliation{\NMSU}
\affiliation{\Oxford}
\affiliation{\Pitt}
\affiliation{\Rutgers}
\affiliation{\SLAC}
\affiliation{\SDSMT}
\affiliation{\Maine}
\affiliation{\Syracuse}
\affiliation{\TelAviv}
\affiliation{\Tennessee}
\affiliation{\UTA}
\affiliation{\Tufts}
\affiliation{\UCL}
\affiliation{\VTech}
\affiliation{\Warwick}
\affiliation{\Yale}

\author{P.~Abratenko} \affiliation{\Tufts}
\author{O.~Alterkait} \affiliation{\Tufts}
\author{D.~Andrade~Aldana} \affiliation{\IIT}
\author{L.~Arellano} \affiliation{\Manchester}
\author{J.~Asaadi} \affiliation{\UTA}
\author{A.~Ashkenazi}\affiliation{\TelAviv}
\author{S.~Balasubramanian}\affiliation{\FNAL}
\author{B.~Baller} \affiliation{\FNAL}
\author{A.~Barnard} \affiliation{\Oxford}
\author{G.~Barr} \affiliation{\Oxford}
% ...

\collaboration{The MicroBooNE Collaboration}
\thanks{microboone\_info@fnal.gov}\noaffiliation

\date{\today}

\begin{abstract}
We present a deep learning-based method for estimating the neutrino energy of charged-current neutrino-argon interactions.
We employ a recurrent neural network (RNN) architecture for neutrino energy estimation in the MicroBooNE experiment, utilizing liquid argon time projection chamber (LArTPC) detector technology.
Traditional energy estimation approaches in LArTPCs, which largely rely on reconstructing and summing visible energies, often experience sizable biases and resolution smearing because of the complex nature of neutrino interactions and the detector response.
The estimation of neutrino energy can be improved after considering the kinematics information of reconstructed final-state particles.
Utilizing kinematic information of reconstructed particles, the deep learning-based approach shows improved resolution and reduced bias for the muon neutrino Monte Carlo simulation sample compared to the traditional approach.
In order to address the common concern about the effectiveness of this method on experimental data, the RNN-based energy estimator is further examined and validated with dedicated data-simulation consistency tests using MicroBooNE data. 
We also assess its potential impact on a neutrino oscillation study after accounting for all statistical and systematic uncertainties and show that it enhances physics sensitivity.
This method has good potential to improve the performance of other physics analyses.
\end{abstract}

\maketitle

\section{Introduction}\label{sec:intro}

Neutrino oscillations, which refer to transitions between different neutrino flavor states along their propagation length, are of strong scientific interest as they unequivocally prove the existence of neutrino mass, which was not predicted by the original Standard Model (SM).
The precise nature of these oscillations is not yet known and is a focus of many experiments~\cite{deSalas:2020pgw}. Some key unanswered questions include the order of neutrino masses, the octant of the mixing angle $\theta_{23}$,  and the presence of leptonic CP violations. The latter question is especially relevant to early universe phenomena such as leptogenesis which has been proposed to explain the observed baryon asymmetry of our universe~\cite{pascoli2007}. In addition, neutrino oscillations serve as a probe for physics beyond the SM such as the hypothetical sterile neutrino states within well-motivated extensions of the SM~\cite{kopp2013sterile}. 

The estimation of neutrino energy $E_{\nu}$ is of crucial importance to experiments studying the phenomenon of neutrino oscillation, since the transition probabilities depend on neutrino energy, in particular on $L/E_{\nu}$, where $L$ is the distance the neutrino has traveled. 
These experiments typically involve the scattering of neutrinos with a broad distribution of energy (i.e., they are not monochromatic) on a fixed nuclear target, which also serves as a detector.
Examples in accelerator experiments can be found in Refs.~\cite{NOvA:2021nfi,t2k:latest,MicroBooNE:2022sdp,DUNE:2020jqi} among others. 
As a result, the neutrino energy is not known a priori on a per-interaction basis and must be 
reconstructed from the interaction itself.

This work focuses on the energy estimation of charged-current (CC) neutrino events since these events are of most interest for oscillation analyses.
The $\nu$-Ar CC neutrino interactions are associated with the weak CC neutrino interaction vertex $\nu_l \to l^- + W^+$ (and $\bar \nu_l \to l^+ + W^-$) and are characterized by the presence of the primary outgoing lepton $l^\pm$.
The energy of the incoming neutrino is transferred to the primary lepton $l^\pm$ and the argon nucleus (through $W^\pm$ bosons).
The latter part of the energy can, in turn, create a number of secondary hadronic particles.
By measuring the energy of the primary lepton and the energies of the resulting hadronic particles one can make inferences about the energy of the incoming neutrino.

For simple CC interactions where the scattering is elastic and on quasi-free nucleons within the nucleus (``charged-current quasi-elastic'' or CCQE), the neutrino energy can be reconstructed from the kinematics of the outgoing charged lepton that is associated with the neutrino, i.e. $\mu$ ($e$) for $\nu_{\mu}$ ($\nu_{e}$)~\cite{mcfarland2011quasi}. For the CCQE interaction $\nu_{\mu} + n \rightarrow {\mu}^{-} + p$, assuming 2-body scattering under the energy and momentum conservation laws,  we have
\begin{equation}\label{eq:qe_energy}
    E_{\nu}^{QE} = \frac{m_{p}^{2} - (m_{n} - E_{b})^{2} - m_{\mu}^{2} + 2(m_{p} - E_{b})E_{\mu}}{2(m_{n} - E_{b} - E_{\mu} + p_{\mu}\cos\theta_{\mu})}, 
\end{equation}
where $m_{p}$, $m_{n}$, and $m_{\mu}$ refer to the mass of the proton, neutron, and outgoing muon, respectively.
$E_{\mu}$ is the muon energy.
$p_{\mu}$ and $\theta_{\mu}$ are the muon momentum and angle with respect to the 
incoming neutrino momentum direction and $E_{b}$ is the binding energy of the proton within the 
nucleus, typically $\sim \mathcal{O}(10)$ MeV for various nuclear targets~\cite{database}.  

However, a significant fraction of neutrino interactions are not quasi-elastic. The nucleons within the nucleus are generally not quasi-free, and the interaction can, therefore, exhibit a strong dependence on the initial nuclear state~\cite{gallagher2011neutrino}. In addition, the incoming neutrino can excite baryon resonances leading to a variety of hadronic final states, typically pions and nucleons~\cite{Mosel:2016cwa}. Moreover, the mechanism of intra-nuclear transport of these final-state hadrons is a research topic itself~\cite{golan2012effects} and introduces an additional layer of uncertainty to the presence as well as the kinematics of final-state hadronic particles. These complications subsequently introduce limitations to the use of Eq.~\eqref{eq:qe_energy}.

A separate strategy to reconstruct $E_\nu$ based on calorimetry is commonly adopted for tracking calorimeters, such as that used in the MicroBooNE experiment~\cite{uboone_detector}. 
The MicroBooNE detector is a liquid argon time projection chamber (LArTPC), which provides mm-scale position resolution, ns-scale timing resolution~\cite{MicroBooNE:2023ldj}, and sub-MeV energy threshold to resolve neutrino interactions in fine detail. 
With the reconstructed information of particles (e.g. type and $4$-momentum) in the final state 
including both the primary and secondary interactions,
the energy of the neutrino is estimated by summing up the
estimated leptonic, hadronic, and nucleon binding energies~\cite{MicroBooNE:2021nxr} based on energy conservation only.
This strategy does not limit itself to the CCQE interactions and is more general.
However, simple summation methods cannot account for energy lost to undetected particles (e.g., particles below threshold, outgoing neutrinos, or neutrons), which leads to missing energy.  Additionally, binding energy corrections can only be performed on average.  Finally, the kinematics of individual particles may be mis-estimated for a variety of reasons, including biases in the reconstruction algorithms, exiting particles, and re-interactions of charged hadrons.  To better estimate neutrino energies, an algorithm needs to be capable of inferring missing energy from the kinematics and topology of the reconstructed particles of individual events.

In this paper, we present a deep learning (DL) based approach to estimate the neutrino energy using both the energy and momentum information for general neutrino-nuclei interactions at the GeV energy scale. Deep learning refers to a class of modern machine learning (ML) techniques that perform deep inference by automatically deriving important representations of the input feature set in a high dimensional space for various tasks such as classification and regression~\cite{lecun2015deep}. This is in contrast to more traditional ML-based approaches that rely on significant human inputs for the feature set.  
Deep learning techniques have been shown to improve performance on a wide array of targeted metrics within high-energy physics~\cite{guest2018deep}, including neutrino physics which is often an early adopter on this front~\cite{radovic2018machine}.
Within neutrino experiments, they have contributed to detector simulation and signal processing~\cite{Yu:2020wxu}, particle identification~\cite{aurisano2016convolutional, MicroBooNE:2018kka}, estimation of the 
interaction vertex~\cite{MicroBooNE:2021ojx}, and reconstruction of energies~\cite{baldi2019improved} and directions~\cite{liu2020deep}, among other tasks.

The DL approach is especially well suited to energy estimation because of the many multi-dimensional inputs, which include the various outgoing particles in the interaction and their particle-flow information, as well as the ability to model non-linear relationships among those inputs in the high-dimensional space. Through training on the simulated event samples, the DL-based approach learns to estimate the neutrino energy considering
correlations of the kinematics of the final-state particles embedded in event generators, which are constrained by both the energy and momentum conservation laws. In this work, we employ recurrent neural networks (RNNs), which have found use in various contexts such as Natural Language Processing (NLP)~\cite{yu2019review}, and are especially suitable for a varying sequence of inputs like the particle flow of a neutrino interaction. 

The DL-based neutrino energy estimator is aiming at a better neutrino energy estimation which can benefit future physics analyses. One example is neutrino oscillation measurement~\cite{MicroBooNE:2022sdp}. 
The sensitivity of the neutrino oscillation analysis depends on the quality of reconstructed neutrino energy spectra and on the ability to resolve various features of these spectra.
To evaluate the performance of the DL-based neutrino energy estimator, we consider two key metrics related to the quality of the reconstructed energy spectra.
These are the bias and resolution, which refer to the mean and root-mean-square (RMS) of the ratio $\frac{(E^{reco} - E^{true})}{E^{true}}$ respectively.
A large energy resolution (RMS) leads to a smearing of features in the oscillated neutrino energy spectrum, reducing sensitivity in measuring neutrino oscillations.
Similarly, a large bias in the energy estimation can skew the oscillation measurement, resulting in a faulty estimate of the oscillation parameters if the bias is not modeled correctly by the simulation.
This can be checked by the model validation procedure shown in \sref{sec:validation}.
Even when the bias in the energy estimation is properly modeled, it can again lead to a reduction of features in the oscillated neutrino energy spectrum, thus decreasing sensitivity.

While the DL-based energy estimator can be shown to outperform traditional energy estimators evaluated with simulation samples, there can be a significant model dependence in mapping from true to reconstructed neutrino energies, given the simulation's task of modeling the complex nuclear physics (i.e. kinematics correlations among final-state particles). 
While such model dependence may not be apparent in evaluations using simulations, it becomes crucial when applying the DL-based energy estimator to experimental data.
In order to mitigate this concern, we perform dedicated model validations with the experimental data from the MicroBooNE experiment to demonstrate that the bias and resolution in the DL-based energy estimator are compatible within the quoted uncertainties of the overall model.
These validations rely on the goodness-of-fit metric~\cite{MicroBooNE:2021nxr}, which is further enhanced by the conditional constraint methodology~\cite{MicroBooNE:2021sfa}.
These dedicated model validations build confidence not only in our overall simulation but also in the DL model, which often suffers from a lack of interpretability because of its black-box nature and sometimes non-intuitive associations among its inputs. 

This paper is organized as follows. In \sref{sec:related_work}, we review various techniques used for neutrino energy estimation and outline the advantages of using the DL-based approach.
In \sref{sec:method}, we introduce the RNN-based DL energy estimator (DL-EE), the preparation of input particle flow information, and the DL model architecture. 
In \sref{sec:training}, we outline the training of the DL-EE on MicroBooNE simulation, including methods to control the resulting output bias of the energy estimator. In \sref{sec:evaluation}, we evaluate the performance of the DL-EE using simulation and show that this technique is able to improve both the resolution and bias of the neutrino energy estimation. In addition, we perform validation of the DL-based energy estimator using experimental data. Furthermore, we study the impact of incorporating the DL estimator on the sensitivity of searching for a sterile neutrino in the $\nu_\mu$ disappearance mode using the MicroBooNE detector~\cite{MicroBooNE:2022sdp} before concluding in \sref{sec:conclusion}.

\section{Review of various neutrino energy estimators}~\label{sec:related_work}

\subsection{The MicroBooNE Experiment}~\label{sec:uboone_exp}

The MicroBooNE detector~\cite{uboone_detector} is a LArTPC consisting 
of $85$ tonnes of liquid argon in the 
active volume, which is a rectangular volume measuring
$10.36$ m in length, $2.32$ m in height, and $2.56$ m in width.
The TPC is placed inside a larger cylindrical 
cryostat which has a total capacity of $170$ tonnes of liquid argon. The MicroBooNE detector sees
the on-axis neutrinos produced from the Booster Neutrino Beam (BNB) at Fermilab National Accelerator Laboratory (FNAL)~\cite{MiniBooNE:2008hfu}. The distance between the MicroBooNE detector
and the BNB target is about 468 m. The neutrinos from the BNB are predominantly ($\sim 93.6\%$) $\nu_{\mu}$ with a mean energy of $0.8$ GeV. 

Ionization electrons are produced by the charged particles from the neutrino interaction traveling through LAr. 
Under an external electric field of $273$ V/cm, these ionization electrons drift horizontally at a constant speed of around
1.1 mm/$\mu$s towards the anode plane.
The anode plane consists of $3$ planes of wires.
The passage of ionization electrons induces bipolar readout signals on two of these planes (``induction planes'') oriented at $\pm 60^\circ$ with respect to the vertical.
The ionized electrons are then collected and induce unipolar signals at the third plane (``collection plane'') oriented vertically.
The wire pitch is $3$ mm.
In addition, a set of $32$ photomultiplier tubes (PMTs) are placed behind the 
wire planes to detect scintillation light from the interaction. The light signal provides a prompt timing 
signal for when the interaction occurred.

Three 2D views of the detector activity can be obtained from the projective LArTPC wire 
channel readouts and drift time. Using tomographic reconstruction algorithms~\cite{Qian:2018qbv}, 
the three 2D views can be combined to create a complete 3D image of the detector activity.

\subsection{Energy Estimators in MicroBooNE}~\label{sec:trad-wc-ee}

In this section, we describe the traditional energy estimation algorithm in MicroBooNE, which
serves as a baseline for the performance comparisons.
In searching for a $\nu_e$ low-energy 
excess~\cite{MicroBooNE:2021tya}, MicroBooNE performed three different analyses, each of which relied on completely different reconstruction paradigms, namely, Wire-Cell~\cite{MicroBooNE:2021nxr}, Pandora~\cite{TheMicroBooNECollaboration:2021cjf}, 
and deep learning~\cite{MicroBooNE:2021pvo}.
Nevertheless, these analyses are based on the same calorimetry neutrino energy estimation 
strategy~\cite{MicroBooNE:2017xvs,Qian:2018qbv,MicroBooNE:2018kka}. 
In the following, we briefly review the neutrino energy reconstruction within the Wire-Cell reconstruction paradigm, upon which the work of this paper is based.

Wire-Cell is a tomography-inspired algorithm that provides a $3$D representation of the interaction 
based on the three $2$D projection measurements from the anode planes~\cite{Qian:2018qbv,MicroBooNE:2020vry}. 
The particles in the final state and their $4$-momenta, including those from 
secondary interactions, are then reconstructed by various downstream algorithms that perform 
TPC-charge/PMT-light matching~\cite{MicroBooNE:2020vry}, trajectory fitting and particle identification, 
and neutrino vertex identification, so that the best description of the interaction (``particle-flow'') 
is produced~\cite{MicroBooNE:2021ojx}.

The traditional Wire-Cell energy estimator used in MicroBooNE is based on the reconstructed particle flow tree 
information and employs a straightforward logic~\cite{MicroBooNE:2021ojx}.
One of three methods is used to reconstruct the deposited energy for each particle: i) range based method for 
track-like particles stopping inside the detector with sufficient length ($>4$cm); ii) summation of $dE/dx$ for 
other track-like particles; iii) scaling of the summed charge for shower-like particles.
The primary final-state charged lepton in CC neutrino interactions
is selected based on reconstructed energy and particle identification information. The visible neutrino energy 
is reconstructed by adding up the single-particle kinematic energies, particle masses, and an averaged 
nucleon binding energy, 8.6 MeV, for each identified proton. More details can be found in
Ref.~\cite{MicroBooNE:2021ojx}.
As discussed in Sec.~\ref{sec:intro}, this particle flow summation method is general and robust against neutrino-argon interaction models, yet it suffers from bias induced by missing energy. This motivates our search for alternative algorithms.

\subsection{Energy Estimation with Convolutional Neural Networks}
The NOvA experiment introduced a DL energy estimator
based on a convolutional neural network (CNN) architecture~\cite{baldi2019improved}.
This energy estimation method utilizes images of neutrino events and tries to predict
the energy of the neutrino from the features present in the image.
The CNN-based approach was shown to give superior energy reconstruction compared to
the traditional energy estimation methods. However, the use of CNN-based
networks is computationally expensive because of large computational
costs associated with the convolutional operations.
In addition, the CNN-based neutrino energy estimator demands an accurate detector simulation
since even tiny pixel-size systematic inaccuracies of the simulation can be captured by the convolutional network, leading to the domain shift problem~\cite{farahani2021brief}.
Recently, a CNN-based approach was applied to 
LArTPC detectors, for the task of shower energy estimation~\cite{carloni2022convolutional}.
As a CNN algorithm, it shares similar trade-offs as the NOvA method.

\subsection{Previous Works on Energy Estimation with Recurrent Neural Networks}
Large computational costs of DL algorithms are associated in part with a large dimensionality (e.g., a large number of pixels) of inputs.
To reduce such computational costs, one approach is to use alternative inputs that have fewer dimensions.
For instance, high-energy physics (HEP) experiments usually run multiple simple and traditional reconstruction 
algorithms over the events (e.g., event type prediction, particle localization, etc.).
The  outputs of these algorithms are relatively low-dimensional and less affected by the potential systematic differences between the simulation and the real experimental data.
Therefore, one can build a robust and less computationally expensive DL energy estimator on top of these outputs.

The application of recurrent neural networks (RNN) to the task of neutrino energy estimation was pioneered by 
the NOvA experiment~\cite{RNN-pioneer}. 
The RNN-based energy estimator consumes information from the reconstructed particles in
each neutrino event and predicts the neutrino energy.
Similar to the CNN-based energy estimation method, the RNN-based method is able to outperform
the traditional energy reconstruction methods~\cite{torbunov2021improving}, but without 
incurring large computational costs.

In this work, we utilize the particle flow output from the state-of-the-art Wire-Cell event reconstruction paradigm~\cite{MicroBooNE:2021ojx} to estimate 
neutrino energy with the RNN. The use of the reconstructed particle flow 
information allows us to suppress the potential difference between the detector 
response simulation and experimental data~\cite{MicroBooNE:2018swd, MicroBooNE:2018vro}

\section{Method Description}\label{sec:method}
\subsection{Energy Estimation with Recurrent Neural Networks}
In this section, we describe how the recurrent neural network energy estimator operates.
The existing Wire-Cell algorithms reconstruct individual particles in the LArTPC
volume~\cite{MicroBooNE:2021ojx}.
These reconstructed particles are hierarchically grouped into a structure
called a particle flow (PF). For each particle in the PF,
we know its starting and ending coordinates, the best estimate of the particle type, 
and rough estimates of the particle's energy and momentum.

For each neutrino interaction event, the \vlne energy estimator aggregates information
from all particles present in the PF structure
in order to make inferences about the
neutrino energy and the energy of the primary outgoing lepton.
The use of the PF information creates unique challenges and opportunities.
First, the number of particles in each event varies, 
depending on the type of the neutrino interaction. Some neutrino interactions can produce just 
a few particles, while others create many of them. Therefore, in order to operate on the PF 
information, the DL model needs to be able to handle inputs of varying lengths.
Second, each particle in the PF information has the same semantic meaning and the same format
of input variables. This structure of inputs can be exploited to create a sample-efficient
deep learning algorithm. In particular, the structure of inputs can be encoded as an
inductive bias of the DL algorithm~\cite{d2021convit}.
Adding such an inductive bias can lead to a model with a smaller number of parameters,
faster training times, and better performance with a limited amount of data.

An RNN is a natural deep learning model candidate to handle the PF information.
An RNN model operates on a sequence of tokens of arbitrary lengths. It treats each token in
the same way, thus ensuring the proper inductive bias. The RNN reads tokens sequentially and maintains a
fixed-size memory state of the past tokens. Once all of the tokens have been consumed, one can use
the resulting memory state to make inferences about the data. For the energy estimation problem, we consider
each particle of the PF to be a separate token, use an RNN to aggregate all particles, and then use another
DL model to predict the energy of the neutrino from the RNN's memory. The detailed
architecture of this energy estimator is described in~\sref{sec:method:model_arch}.

There is another, more recent, candidate DL model that can be used to work with the PF data -- the transformer model~\cite{vaswani2017attention}.
In many domains, transformer-based models achieve
state-of-the-art performance~\cite{dosovitskiy2020image}.
However, in order to achieve outstanding performance they require large amounts of training data, which is not available at \uboone.
Therefore, we have chosen to use the RNN-based model in this work.

\subsection{Model Architecture}~\label{sec:method:model_arch}
The \vlne energy estimator is designed to predict both the energy of the neutrino and the energy of the primary final-state
lepton in a neutrino interaction event. To make such predictions, it relies on reconstructed particle information
available for each neutrino event (PF information). As inputs, the \vlne estimator extracts the
following quantities from each particle:
i) particle track starting and ending coordinates; ii) estimated particle momentum and energy; iii) 
estimated particle type. These quantities are reconstructed by the upstream Wire-Cell 3D pattern 
recognition algorithms~\cite{MicroBooNE:2021ojx}. Besides the particle-level information, the \vlne energy 
estimator incorporates the information about the entire event such as:
i) a flag indicating whether the event  is fully (FC) or partially contained (PC) in the detector~\cite{MicroBooNE:2021nxr} volume;
ii) a prediction on whether the event is a $\nu_\mu$ CC or a $\nu_e$ CC event.
The FC events are defined as those that have the reconstructed TPC activity fully contained within the fiducial volume (3 cm inside the effective TPC boundary~\cite{MicroBooNE:2021zul}).

\begin{figure*}
  \centering
  \includegraphics[]{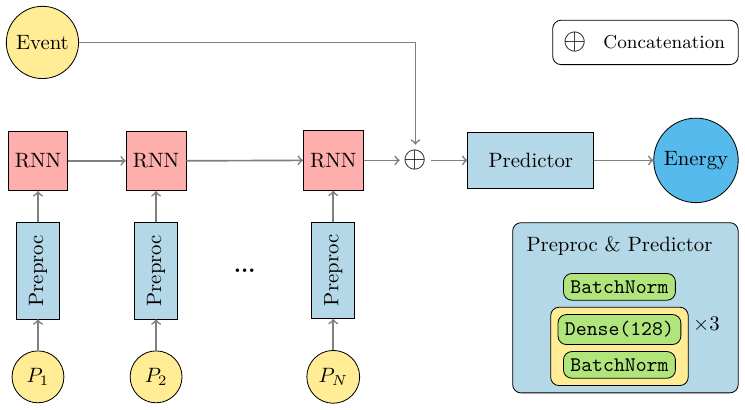}
  \caption{
    The schematic representation of the energy estimator architecture. The yellow circle
    labeled \textit{Event} denotes the event-level inputs. The circles labeled
    \textit{$P_1$} \ldots \textit{$P_N$} represent input variables coming from
    the reconstructed particles ($1 \ldots N$) in the event. \textit{RNN} is
    a recurrent neural network cell. \textit{Preproc} is a particle-wise
    feed-forward neural network used to perform the feature extraction from the
    particle-level variables. \textit{Predictor} is another feed-forward
    neural network that predicts neutrino and lepton energies.
  }
  \label{fig:rnn_ee_arch}
\end{figure*}

The architecture of the \vlne is shown in \fref{fig:rnn_ee_arch}. We chose
a long short-term memory (LSTM)~\cite{van2020review} neural network cell as a recurrent 
neural network model since it is rather stable to train.
Before feeding the information from each particle into the LSTM cell, we perform a
feature extraction step with the help of a simple fully connected network~\cite{zhang2023dive} (depicted as the \texttt{Preproc}
branch in \fref{fig:rnn_ee_arch}). After the LSTM cell has finished processing
particles in an event, the memory state of the LSTM (vector of 32 features) is concatenated with the event level
information and fed through another fully connected network (depicted as \texttt{Predictor}),
which is responsible for the actual prediction of the neutrino energy and the energy
of the primary lepton in the event.

\section{DL Model Training}\label{sec:training}
The \vlne energy estimator is trained on a simulated dataset made of true-$\nu_e$ and true-$\nu_\mu$ CC events.
Both FC and PC events are included in the training sample.
In this section, we show the performance on $\nu_\mu$CC events, where a large amount of data 
events are available to perform dedicated model validations, as described in \sref{sec:evaluation}.

\subsection{Training Dataset}
\label{sec:train:dataset}

In the MicroBooNE experiment, the simulated BNB neutrino flux~\cite{MiniBooNE:2008hfu} is provided to the event generator 
Genie~\cite{Andreopoulos:2009rq,GENIE:2021npt} to generate neutrino-argon interactions. 
Genie v3.0.6, G18\_10a\_02\_11a, was used, which includes improvements on the use of the 
Valencia model~\cite{Nieves:2011yp,Nieves:2004wx,Gran:2013kda} for the local Fermi gas nucleon 
momentum distributions, improvements in the CCQE and CC two-particles-two-holes (CC2p2h) interactions, and
improvements in the treatments of final state interaction
(FSI) and pion production with respect to earlier versions. In addition to the default configuration, 
the parameters governing the CCQE and CC2p2h models are adjusted according to the T2K CC0$\pi$ cross-section 
results~\cite{T2K:2016jor} to form the ``MicroBooNE Tune'' model~\cite{MicroBooNE:2021ccs}. The resulting 
final-state particles of each Monte Carlo (MC) simulated event are processed using the LArSoft~\cite{Snider:2017wjd} software 
framework, which is a toolkit to perform simulation, reconstruction, and analysis of LArTPC data. 
The final state particles are propagated  through the detector using the Geant4 toolkit~\cite{GEANT4:2002zbu} 
v4\_10\_3\_03c. The resulting energy depositions are further processed by dedicated detector simulation programs 
taking into account detector effects to simulate the ionization charge and scintillation light signals after 
considering the space charge effect~\cite{Abratenko:2020bbx,Adams:2019qrr}.

The position and number of ionization electrons modified by space charge and recombination effects are
ported to the TPC detector simulation~\cite{MicroBooNE:2018swd, MicroBooNE:2018vro}, which takes into account the charge 
transportation and diffusion~\cite{Li:2015rqa}. The induced currents on the wires are simulated by 
convolving the ionization charge distribution at the wire plane with the position-dependent (at 1/10th of 
the wire pitch resolution) field response function as well as the electronics response function.
The optical detector simulation models the light emitted by charged particles interacting with the 
detector and produces signals in photomultiplier tubes. 

The simulated neutrino interactions are further merged with a dedicated data stream
which is collected in a period when there is no neutrino beam, ensuring faithful modeling of
cosmic-ray backgrounds and detector noise.
At the same time, this choice limits the number of available simulation 
events due to a finite number of cosmic-only data events.
We split the simulated neutrino dataset into training/test partitions following a previously used \uboone procedure~\cite{collaboration2022search}.
Table~\ref{tab:trai_test_data_size} summarizes 
the final sample sizes used in the training and testing. 
\begin{table}
    \centering
    \caption{Number of neutrino CC events in the training and test datasets obtained from the MicroBooNE simulation.}
    \label{tab:trai_test_data_size}
    \begin{tabular}{|l|r|r|}
        \hline
        & \textbf{Train} & \textbf{Test} \\
        \hline
        $\nu_{\mu}$ FC & 80,256 & 134,707 \\
        $\nu_{\mu}$ PC & 156,628 & 303,422 \\
        $\nu_{e}$  FC & 93,289 & 112,822 \\
        $\nu_{e}$  PC & 66,792 & 68,061 \\
        \hline
    \end{tabular}
\end{table}

\subsection{Initial Training}~\label{sec:train_initial}
Initially, we trained the \vlne energy estimator~(\fref{fig:rnn_ee_arch}) on a dataset described in 
Table~\ref{tab:trai_test_data_size} including all 4 samples (\{$\nu_{\mu}$, $\nu_{e}$\} $\times$ \{FC, PC\}).
The model was trained to predict both the energy of the neutrino and the energy of the primary lepton 
in the event. As a loss function for each target (neutrino and primary lepton), we used a mean absolute percentage error:
\begin{equation}
  \label{eq:loss}
  \mathcal{L}_{\nu, \text{lep}} = 100 \cdot \left|
    \frac{E^{\text{reco}} - E^{\text{true}}}{E^{\text{true}}}
  \right|
\end{equation}
where $E^{\text{true}}$ is the true energy of the particle, and $E^{\text{reco}}$ is its predicted energy.

The total loss function is a sum of losses for each target, i.e. $\mathcal{L} = \mathcal{L}_{\nu} + \mathcal{L}_{\text{lep}}$.
We have found this loss function to perform much better than the traditional
regression losses (mean squared error and mean absolute error~\cite{zhang2023dive}). The training was performed for 200 epochs (complete passes through the training dataset) with a \texttt{ReduceLROnPlateau} learning rate scheduler~\cite{ReduceLROnPlateau}
allowing the energy estimator to converge. The complete training details can be found
in \aref{sec:app:train_details}.

The sequential nature of the RNN network may make the energy estimator dependent on the particular ordering of the particles used in the training.
To ensure that the energy estimator performance does not depend on the particle ordering, we have implemented particle order randomization as a data augmentation strategy.

\begin{figure}[h]
  \centering
  \includegraphics[width=0.45\textwidth]{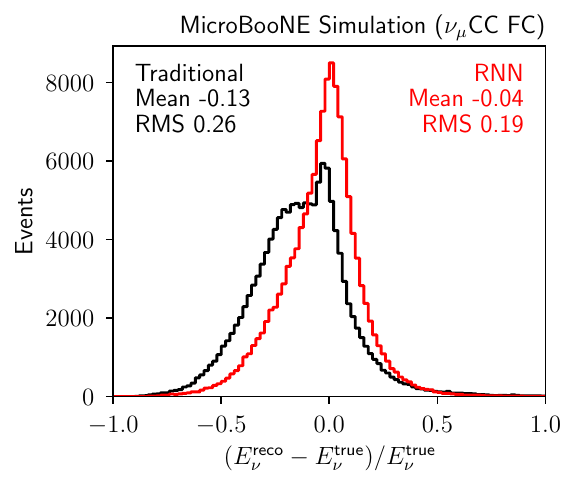}
  \caption{
    $\nu_\mu$ energy resolution histograms for the traditional \uboone
    energy estimator (black) and the initial training of the \vlne energy
    estimator (red) in the Fully Contained (FC) CC sample.
  }
  \label{fig:vlne_v2_numu_fc_resolution_total}
\end{figure}

To assess the performance of the energy estimator, one can consider its energy resolution defined as a ratio of $(E^{\text{reco}} - E^{\text{true}})/E^{\text{true}}$.
\Fref{fig:vlne_v2_numu_fc_resolution_total} shows distributions of the $\nu_\mu$ CC energy resolutions of the FC events.
The distribution of the \vlne energy resolution (shown in red) has a smaller width compared to the traditional energy estimator (shown in black).
This indicates that the \vlne energy estimator is able to better predict the true neutrino energy $E^{\text{true}}$.

For a more quantitative assessment of the performance, one can consider two characteristics of the energy resolution -- its mean and RMS values
\begin{align}
    \text{Mean} &:= \mathbb{E} \left[
        \frac{E^{\text{reco}} - E^{\text{true}}}{E^{\text{true}}}
    \right] \\
    \text{RMS} &:= \sqrt{
        \mathbb{E} \left[
            \left(
                \frac{E^{\text{reco}} - E^{\text{true}}}{E^{\text{true}}}
            \right)^2
        \right]
    }
\end{align}
where $\mathbb{E}$ is an expectation over the test sample.
Generally, smaller RMS values correspond to better energy estimators. The mean value indicates the overall bias of the energy estimator.
According to \fref{fig:vlne_v2_numu_fc_resolution_total}, the \vlne energy estimator (RMS $\sim 0.19$) achieves about $26 \%$ improvement compared to the traditional \uboone energy reconstruction method (RMS $\sim 0.26$) in the test sample.

\begin{figure}[h]
  \centering
  \includegraphics[width=0.45\textwidth]{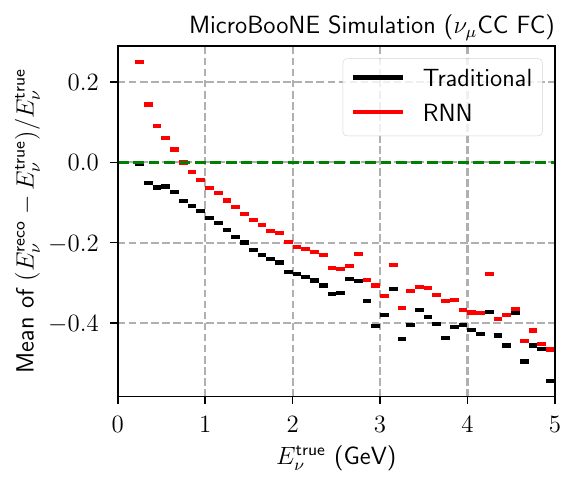}
  \caption{
    $\nu_\mu$ energy bias for the traditional \uboone
    energy estimator (black) and the initial training of the \vlne energy
    estimator (red) in the Fully Contained (FC) CC sample.
  }
  \label{fig:vlne_v2_numu_fc_binstat_total_mean}
\end{figure}

Another metric that is commonly considered in HEP experiments is the bias of
the energy estimator. To determine a bias of the energy estimator, a binned statistics
plot is made, where the x-axis represents the true energy and the y-axis shows
a mean of the energy resolution histogram constructed for each bin. 
\Fref{fig:vlne_v2_numu_fc_binstat_total_mean}
shows the bias plot for the \vlne energy estimator (red) and the traditional \uboone energy estimator (black).
Any deviation from zero represents the bias of the energy estimator in a particular true energy bin.

According to \fref{fig:vlne_v2_numu_fc_binstat_total_mean}, the \vlne energy estimator 
has a smaller bias compared to the traditional \uboone energy estimator for energies above
$\sim \SI{0.6}{GeV}$. However, at lower energies the \vlne energy estimator quickly
acquires a rather significant bias.
Such a large bias is not ideal,
since it may lead to a reduction in physics sensitivities. 

\subsection{Reducing Bias}

In this section, we develop a mitigation strategy for the large bias
of the \vlne energy estimator. Before developing such a strategy, it
is instructive to consider the reasons for the appearance of
the bias. We believe there are two main sources of bias, one is related to physics, and the other is related to ML.

From the physics point of view, as the energy of the neutrino gets higher, a larger
fraction of this energy becomes invisible in neutrino-argon interactions.
This happens due to the increased production of various mesons at higher energies.
Meson decays have a common byproduct -- neutrinos, which easily escape the detector carrying some fraction of the energy of the original interaction away.
Therefore, one may expect to see an increasingly negative (missing energy) bias at
high energies for all estimators.
\Fref{fig:vlne_v2_numu_fc_binstat_total_mean} shows
that both energy estimators acquire large negative biases at high energies.

\begin{figure}[h]
  \centering
  \includegraphics[width=0.45\textwidth]{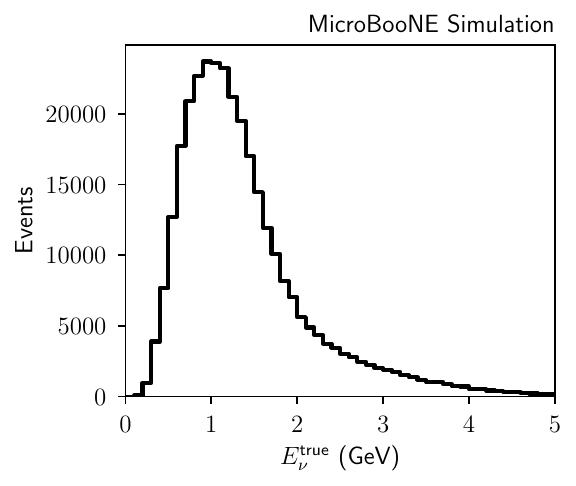}
  \caption{
    True neutrino energy distribution in the training sample.
  }
  \label{fig:energy_hist_numu+nue_fcpc}
\end{figure}

From the ML side, the peaked nature of the energy distribution
in the training sample (\fref{fig:energy_hist_numu+nue_fcpc}) can also contribute to the bias.
Since the target energy distribution has high population around the $\SI{1}{GeV}$ peak, energy estimators will prioritize correctly reconstructing the energy of neutrinos around the peak.
Moreover, on average, an energy estimator can increase the overall accuracy
of predicted energies by slightly pushing all the energies towards the peak.
That is, the energy of neutrinos to the left of the peak will be pushed up
(positive bias), and the energy of neutrinos to the right of the peak will
be pushed down (negative bias). This prediction is consistent with the
observed behavior of the bias in \fref{fig:vlne_v2_numu_fc_binstat_total_mean}.

The proper way to improve the bias, stemming from the peaked nature of the distribution,
is to re-simulate the training sample with a flat true neutrino energy distribution.
Since we are limited by the total amount of 
simulation events available, we decided instead to use the standard ML approaches to 
deal with imbalanced data. In particular, we applied event reweighting to flatten the true
neutrino energy distribution. The new event weights were constructed using the following procedure.

A histogram of the true neutrino energy $N_i$ was made in a range of
$0 - \SI{5}{GeV}$ with 50 bins. A weight histogram $W_i$ was constructed, so that 
each bin's height is proportional to the the inverse of the true neutrino energy 
bin height, i.e. $W_i \propto 1/N_i$.
In training the \vlne energy estimator, the weight for each event was determined
from the corresponding bin of the weight histogram $W_i$.
To preserve the scale of the loss function, the weights were normalized
to add up to unity.

\begin{figure}[h]
  \centering
  \includegraphics[width=0.45\textwidth]{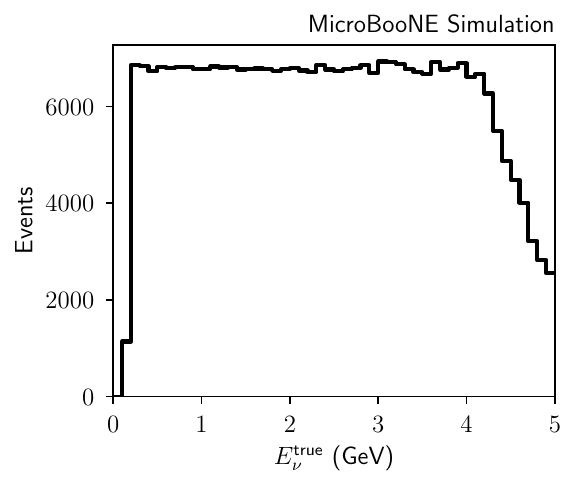}
  \caption{
    Reweighted true neutrino energy distribution in the training sample.
  }
  \label{fig:energy_hist_numu+nue_fcpc_flat}
\end{figure}

The weight construction procedure described above guarantees that the true neutrino energy
distribution is approximately flat. However, the tails of the distribution in 
\fref{fig:energy_hist_numu+nue_fcpc} will have to acquire quite large weights,
since there are too few events in those tails. Giving large weights
to a few events will result in severe overfitting in the model training. In order to reduce 
the magnitude of overfitting,
we clipped the maximum weight value, ensuring the ratio of the maximum weight to the minimum weight does not exceed 50.
\Fref{fig:energy_hist_numu+nue_fcpc_flat} shows the true neutrino energy spectrum
of the training sample after the reweighting.

\begin{figure}[h]
  \centering
  \includegraphics[width=0.45\textwidth]{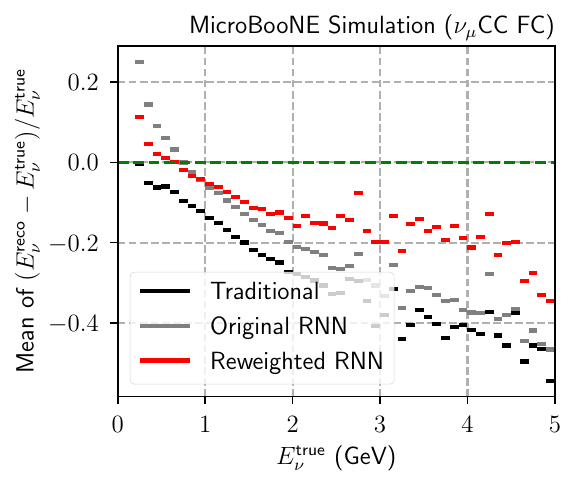}
  \caption{
    $\nu_\mu$ energy resolution bias for the traditional \uboone
    energy estimator (black) and the reweighted training of the \vlne energy
    estimator (red) in the Fully Contained (FC) CC sample.
  }
  \label{fig:vlne_v3_numu_fc_binstat_total_mean}
\end{figure}

\Fref{fig:vlne_v3_numu_fc_binstat_total_mean} shows the $\nu_\mu$ neutrino energy
bias after retraining with the flat weights. It demonstrates that the use of the
reweighted training sample improves the bias of the energy estimator.

\section{DL Model Evaluation}\label{sec:evaluation}

\subsection{Performance Metrics in Testing Simulation Sample}

\begin{figure}[h]
  \centering
  \includegraphics[width=0.45\textwidth]{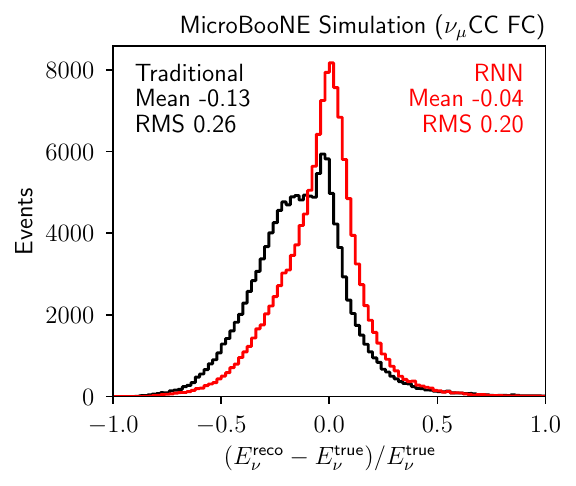}
  \caption{
    $\nu_\mu$ energy resolution histograms for the traditional \uboone
    energy estimator (black) and the reweighted training of the \vlne energy
    estimator (red) on the Fully Contained (FC) CC sample.
    The ``shoulder'' at the lower reconstructed energy comes from a combined effect of missing hadronic energy and the biased recombination model.
  }
  \label{fig:vlne_v3_numu_fc_resolution_total}
\end{figure}

In the previous section, we showed the basic $\nu_\mu$ energy reconstruction performance plots for 
the \vlne energy estimator in the test sample.
We now perform a more detailed evaluation of 
the \vlne energy estimator. \Fref{fig:vlne_v3_numu_fc_resolution_total} demonstrates
the $\nu_\mu$ neutrino energy resolution after retraining with the flat weights.
The use of the reweighted training sample produces a small degradation of the \vlne energy resolution (RMS increases from $\sim 19\%$ to $\sim 20\%$).

We believe that the reduction of the bias of the energy estimator, brought by the reweighting, outweighs the associated small degradation of the RMS value.
This is partly because the reduction of the bias makes the energy estimator more agnostic to the choice of the neutrino energy spectrum.
For instance, the original behavior of the bias stems from the BNB energy spectrum having a peak around $\SI{0.8}{GeV}$.
However, \uboone can also study the NuMI neutrino beam~\cite{Aliaga:2016oaz} with higher average energies.
Having an unbiased energy estimator would allow us to expect better transferability of its performance between different neutrino beams.

Since the reweighted training results in a better energy estimator overall, we will use the reweighted version in the subsequent analysis.
Likewise, each time we refer to the RNN energy estimator below, we will imply its reweighted version.

\begin{figure}[h]
  \centering
  \includegraphics[width=0.45\textwidth]{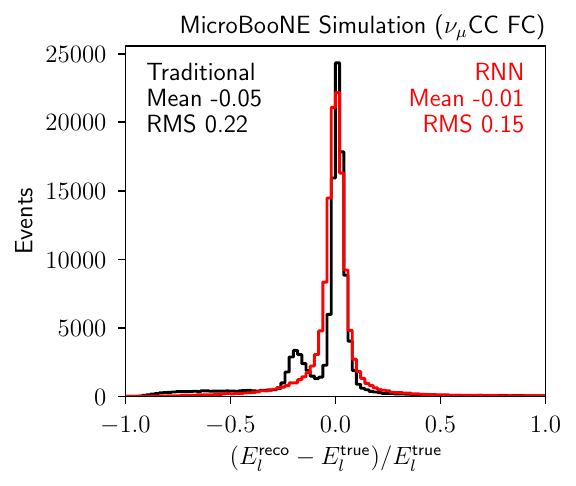}
  \caption{
    Primary $\mu$ energy resolution histograms for the traditional \uboone
    energy estimator (black) and the reweighted training of the \vlne energy
    estimator (red) on the Fully Contained (FC) CC sample.
  }
  \label{fig:vlne_v3_numu_fc_resolution_primary}
\end{figure}

Apart from predicting the $\nu_\mu$ energy, the \vlne energy estimator is also capable of
predicting the energy of the primary lepton.
\Fref{fig:vlne_v3_numu_fc_resolution_primary}
shows energy resolution histograms for the energy reconstruction of the primary outgoing $\mu$ in the $\nu_\mu$ CC events.
The traditional energy estimator exhibits two peaks in the shape of the muon energy resolution.
We discuss the nature of the peaks later in this section.
The \vlne energy estimator has a well-behaved energy resolution histogram with a single peak.
Overall, the \vlne energy estimator displays better energy resolution in terms of
RMS ($15 \%$ vs. $22 \%$).

The unusual shape of the energy resolution of the traditional estimator is traced back
to reconstruction errors.
The central (main) peak has a small bias and corresponds to correctly identified muons.
The energies of these muons are mostly reconstructed from range information.
The resolution of these events is, therefore, good. 
The second peak with a negative bias is traced back to muons
for which the energy was reconstructed by integrating the energy loss per unit length $dE/dx$. 
Biases in the detector modeling, including the inaccuracies in charge recombination, contribute to this offset. This is evident as a second peak in \fref{fig:vlne_v3_numu_fc_resolution_primary} and an enhanced left ``shoulder'' in \fref{fig:vlne_v3_numu_fc_resolution_total}.
The same recombination model was used for reconstructing the simulation, and dedicated validation tests have demonstrated its consistency with the data~\cite{MicroBooNE:2021nxr, MicroBooNE:2021sfa}.

As shown in \fref{fig:vlne_v3_numu_fc_resolution_primary},
the peak of the \vlne's muon energy resolution is shorter and wider than the first (main) peak
of the traditional energy estimator. The degradation of the energy resolution in the main
peak is likely a result of the inability of the \vlne energy estimator to differentiate
between various types of the imperfect event reconstruction.
Because the \vlne energy estimator uses reconstructed particle flow information, it inherits all the shortcomings of the particle flow reconstruction.
In addition, unlike the traditional energy estimator, we do not provide any explicit indicators
of the reconstruction quality to the \vlne model. Therefore, even for the properly reconstructed muons,
we may anticipate a degradation of the performance of the \vlne method compared to the traditional method.

In principle, in the absence of explicit indicators of the reconstruction failures,
a DL model could try to infer them from some other features
(e.g. incorrect event topology).
However, it is unclear what fraction of information about the reconstruction quality can be inferred from the existing features.
Moreover, very large datasets are required for DL models to make such complex inferences. Therefore, the relatively small size of our training dataset may contribute
to the degradation of the quality of the main muon peak.

\begin{figure}[h]
  \centering
  \includegraphics[width=0.45\textwidth]{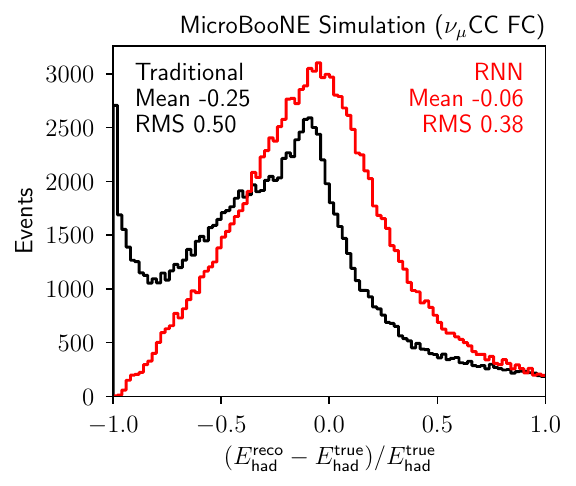}
  \caption{
    Hadronic energy resolution histograms for the traditional \uboone
    energy estimator (black) and the reweighted training of the \vlne energy
    estimator (red) on the Fully Contained (FC) CC sample.
  }
  \label{fig:vlne_v3_numu_fc_resolution_secondary}
\end{figure}

While the \vlne energy estimator does not directly predict the hadronic part of the neutrino energy, it
can be trivially inferred by subtracting the energy of the primary muon from the total
neutrino energy. \Fref{fig:vlne_v3_numu_fc_resolution_secondary} shows the hadronic
energy resolution histograms for the reweighted \vlne and the traditional \uboone energy estimators.
The \vlne hadronic energy resolution exhibits more Gaussian-like behavior and has a much better
RMS ($\sim38\%$ vs. $\sim 50\%$).

In this section, we have studied the performance of the \vlne energy estimator on the sample
of fully contained $\nu_\mu$ CC events.
The \vlne estimator is also capable of predicting the energy of partially contained $\nu_\mu$ CC events, and the energy of $\nu_e$ CC events.
The corresponding performance evaluations are provided in the appendices.
In particular, 
the \vlne energy estimator performance for the $\nu_e$ events is discussed in \aref{sec:app_nue}.
Likewise, \aref{sec:app_pc_events} demonstrates the performance of the \vlne energy estimator 
on the partially contained events.
Finally, \aref{sec:app_resolution_vs_energy} discusses the behavior of the \vlne energy
resolution for various ranges of true neutrino energy.

\subsection{Model Validation with Experimental Data}\label{sec:validation}

When developing new ML algorithms for HEP experiments, it should always be kept in mind that the algorithms are trained on a simulated sample, but eventually are applied 
to experimental data.
The simulation 
data in HEP experiments commonly possess systematic differences from the real experimental data. For example,
as briefly reviewed in \sref{sec:intro}, the response of argon nuclei to a neutrino probe 
depends on the complex nuclear structure and quantum chromodynamics in the non-perturbative region,
which is at the frontiers of nuclear physics research. In addition, understanding
LArTPCs' response and calibration have considerable room for improvement.  Because of these differences 
between data and simulation, it is possible that ML algorithms have systematically different 
performance when applied to real experimental data than those evaluated through simulations. 

HEP communities have developed a strict scheme of estimating systematic uncertainties to 
evaluate and quantify the differences between data and simulation.
Different sources of systematic uncertainties serve as effective degrees of freedom in describing the differences between prediction (e.g., simulation) and data.
Instead of requiring that the simulation
faithfully reproduce every feature in the data, the HEP community requires that the differences between the data and 
simulation are within the quoted systematic uncertainties. In other words, the simulation is required 
to be compatible with data within its quoted uncertainties. 

Since the \vlne energy estimator combines the calorimetry information,
which is sensitive to detector response modeling,
and the kinematics information, which is sensitive to the complex neutrino-argon interaction, 
we expect differences in performance between data and simulation.
We utilize a MicroBooNE data set to demonstrate the compatibility between simulation that is enhanced with the \vlne energy estimator and data.
This data set was collected from February 2016 to July 2018 
corresponding to an exposure of 6.369$\times$10$^{20}$ protons on target from the BNB
at FNAL that was used to search for a low-energy $\nu_e$ excess~\cite{MicroBooNE:2021nxr}. 
As elaborated in Ref.~\cite{MicroBooNE:2021nxr}, the sources of systematic uncertainties associated 
with the simulation include i) neutrino flux, ii) neutrino-argon interaction cross sections, iii) detector effects, 
and iv) statistical uncertainties due to a finite number of simulation events with unequal weights.

Following previous work~\cite{MicroBooNE:2021sfa,MicroBooNE:2021nxr}, the primary tools to test the compatibility between data and simulation (i.e. model validation)  are based on goodness-of-fit (GoF) tests, which allow one to quantify the comparison of data and predictions into a single number for evaluation.
As detailed in Section~V of~\cite{MicroBooNE:2021nxr}, the term ``simulation'' within the context of MicroBooNE refers to the \OverallModel. This model consists of various components, including the neutrino beam flux model, the neutrino-argon interaction model, the detector model, and the reconstruction algorithms, among others. Therefore, changes in the energy estimator impact the consistency between data and simulation.

While the total GoF
test is essential, it may hide some problems of the model when some model uncertainties are overestimated. 
Additionally, since the missing hadronic energy $E^{\rm miss}$ cannot be directly
measured, an event generator (or a neutrino-nucleus interaction model) is often required to describe $E^{\rm miss}$ accurately in order to ensure a correct mapping from $E_{\nu}^{\rm reco}$ to $E_{\nu}^{\rm true}$.
This mapping is crucial as neutrino oscillation measurements rely on estimations of $E_{\nu}^{\rm true}$. However, modeling $E^{\rm miss}$ remains a challenging theoretical problem, particularly for heavy nuclei such as argon where final state interactions can produce a variety of complex final states that contain significantly different amounts of missing energy, even for a simple quasi-elastic interaction.
In order to mitigate these shortcomings, we implement a conditional constraining procedure.
In Refs.~\cite{MicroBooNE:2021sfa,MicroBooNE:2021nxr},
this procedure was used to validate the modeling of missing hadronic energy and its associated uncertainties.
The validation was performed by comparing the reconstructed hadronic energy distribution between data and an MC prediction after constraining 
the reconstructed muon kinematic distributions (i.e. energy and polar angle) to those of data. 
The validation of the mapping between the true and the reconstructed neutrino energy enables
the measurement of neutrino-energy dependent total and differential cross 
sections~\cite{MicroBooNE:2021sfa, MicroBooNE:2023foc} as well as searches for 
sterile-neutrino-induced oscillations~\cite{MicroBooNE:2022sdp}. 

In the following, we briefly review the model validation procedure, which is based on a covariance matrix 
formalism in constructing the $\chi^2$ test statistic:
\begin{equation}
    \chi^2 = \left(M-P\right)^T\times Cov_{\text{full}}^{-1}\left(M,P\right) \times \left(M-P\right),
    \label{eq:test-stat-chi2}
\end{equation}
where $M$ and $P$ are vectors of measurement and prediction, respectively. The $Cov\left(M,P\right)$ is the full covariance matrix:
\begin{dmath}\label{eq:test-stat-1}
    Cov_{\text{full}} = Cov^{\text{stat}}_{\text{stat}} + Cov^{\text{sys}}_{\text{MC stat}} + Cov^{\text{sys}}_{\text{xs}} + Cov^{\text{sys}}_{\text{flux}} + Cov^{\text{sys}}_{\text{det}} + Cov^{\text{sys}}_{\text{add}}. 
\end{dmath}
The $Cov^{\text{stat}}_{\text{stat}}$, $Cov^{\text{sys}}_{\text{MC stat}}$,  $Cov^{\text{sys}}_{\text{xs}}$, $Cov^{\text{sys}}_{\text{flux}}$, $Cov^{\text{sys}}_{\text{det}}$, and $Cov^{\text{sys}}_{\text{add}}$
terms represent the statistical uncertainties of the data sample, the statistical uncertainties corresponding to finite statistics
in simulation, systematic uncertainties in
cross section modeling, systematic uncertainties from the modeling of the neutrino flux, systematic uncertainties from detector response modeling, and additional systematic uncertainties
associated with estimating background events from outside the cryostat, respectively.

\begin{figure}[h]
  \centering
  \includegraphics[width=0.5\textwidth]{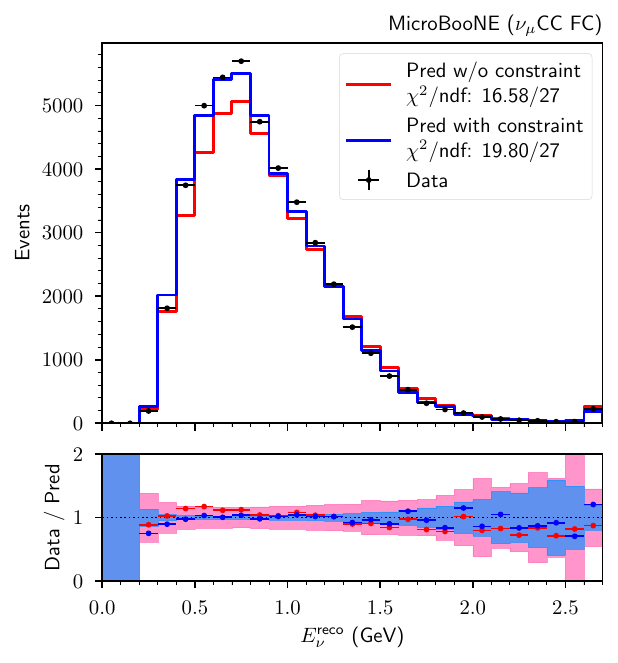}
  \caption{
  Top: Distribution of the selected FC $\nu_{\mu}$CC events as a function of the reconstructed neutrino energy.
  The MC prediction after applying constraints on muon kinematics ($E_{\mu}^{\rm reco}$ and $cos_{\theta}^{\rm reco}$) and hadronic energy ($E_{had}^{\rm reco}$) is shown in blue,
  and before applying in red.
  The last bin represents all events with $E_{\nu}^{\rm reco} >$ 2.6~GeV.
  Bottom: The blue (red) points represent the ratio between data and the MC prediction with (without) constraint, and the bands with same colors depict the $\pm$1$\sigma$ of the total uncertainty (statistical and systematic) of the MC central prediction.}
  \label{fig:gof_e_nu_fc_wi}
\end{figure}

The GoF evaluation is performed to test the compatibility between the data and the \OverallModel
using Eq.~\eqref{eq:test-stat-1}. The $\chi^2$ value can be used to perform a GoF test and to deduce a p-value 
by comparing to the $\chi^2$ distribution with the associated number of degrees of freedom ($ndf$), which is the 
total number of bins used in the measurement.
For example, \fref{fig:gof_e_nu_fc_wi} demonstrates a data-MC comparison of the selected FC $\nu_{\mu}$ CC events as a function of reconstructed neutrino energy.
The p-value is above 0.05, which is a pre-defined threshold for each GoF test.
We should note that the MicroBooNE \OverallModel contains many conservative
systematic uncertainties (e.g. cross section uncertainties). Therefore, the reduced $\chi^2$ values, which are the ratios between 
$\chi^2$ and number of degrees of freedom, are generally low suggesting that 
the \OverallModel describes the data well within its uncertainties.

The conservative estimation of systematic uncertainties aims to determine uncertainties that are large enough to cover all reasonable systematic differences between simulated and real data.
However, the resulting uncertainties may be overestimated, potentially obscuring deficiencies in the \OverallModel.
To address this shortcoming, the global goodness-of-fit test can be enhanced to study different parts of the \OverallModel using the conditional covariance matrix formalism~\cite{cond_cov}.

For example, consider two quantities (channels) $X$ and $Y$ with the goal of assessing data-simulation differences between their simulated predictions $(X_{\text{MC}}, Y_{\text{MC}})$ and the actual measurements $(X_{\text{Data}}, Y_{\text{Data}})$.
These quantities could correspond to any measurements, such as the reconstructed neutrino energy, muon energy, muon angle, etc.
One can perform a direct GoF test by comparing $X_{\text{MC}}$ to $X_{\text{Data}}$ (and similarly for $Y$), but such a test may suffer from the overestimation of systematic uncertainties.
Additionally, if one can simulate conditional probabilities of the form $P(X | Y)$, then the difference between $Y_{\text{MC}}$ and $Y_{\text{Data}}$ can be used to refine $X_{\text{MC}}$ and constrain the magnitudes of its systematic uncertainties.
This allows us to create a more stringent GoF test. The procedure of refining $X_{\text{MC}}$ from $Y$ is referred to as constraining $X$ on $Y$
using the conditional constraining method.

\Fref{fig:gof_e_nu_fc_wi} illustrates data-simulation comparisons of the \vlne neutrino energy for $\nu_\mu$ CC FC events.
The red curve shows the direct prediction of the neutrino energy with the \vlne energy estimator.
The blue curve is a prediction of the neutrino energy after constraining on muon kinematics (reconstructed muon energy and angle) and hadronic energy simultaneously.
The constraining procedure significantly reduces the magnitudes of the systematic uncertainties (shaded areas in the ratio plot) and provides a more precise GoF test.

Following this methodology, we have run a comprehensive and systematic set of \uboone validation tests on 
the \OverallModel that has been enhanced by the \vlne energy estimator using CC $\nu_\mu$ interactions. 
We should note that the CC $\nu_e$ interactions are limited by the data statistics at O(100), 
which is not sufficient to 
perform precision tests on the models. Table~\ref{tab:vlne_v3_data_mc_valiadtion} summarizes the results of these 
model validation tests, with the $p$-values above $0.05$ indicating a successful test.
The \vlne energy estimator passes all the tests successfully, suggesting that the 
difference between data and simulation are within the quoted uncertainties. 

\begin{table}[h]
\centering
\begin{tabular}{|l|r|r|}
\hline
\multicolumn{1}{|c|}{\multirow{2}{*}{Test}} & \multicolumn{2}{c|}{$p$-value} \\
\cline{2-3} & Traditional & \vlne \\
\hline
$P(E_\mu)$ & 0.89 & 0.92 \\
$P(E_{\text{had}})$ & 1.00 & 0.99 \\
$P(E_\nu)$ & 1.00 & 0.97 \\
\hline
$P(E_\mu^{\text{PC}} | E_\mu^{\text{FC}})$ & 0.95 & 0.70 \\
$P(E_{\text{had}}^{\text{PC}} | E_{\text{had}}^{\text{FC}})$ & 1.00 & 0.96 \\
$P(E_\nu^{\text{PC}} | E_\nu^{\text{FC}})$ & 1.00 & 0.84 \\
\hline
$P(E_\mu | \cos\theta_\mu)$ & 0.45 & 0.50 \\
$P(E_{\text{had}} | \cos\theta_\mu)$ & 1.00 & 0.99 \\
$P(E_\nu | \cos\theta_\mu)$ & 1.00 & 0.97 \\
\hline
$P(E_{\text{had}} | E_\mu)$ & 1.00 & 0.97 \\
$P(E_\nu | E_\mu)$ & 1.00 & 0.99 \\
\hline
$P(E_{\text{had}} | E_\mu, \cos\theta_\mu)$ & 1.00 & 0.99 \\
$P(E_\nu | E_\mu, \cos\theta_\mu)$ & 1.00 & 1.00 \\
\hline
$P(E_\nu | E_\mu, \cos\theta_\mu, E_{\text{had}})$ & 1.00 & 0.97 \\
\hline
\end{tabular}
\caption{
    Data vs MC validation results of the reweighted \vlne energy estimator.
    The first column of the table shows the label of the statistical
    test that was performed. The second and third columns indicate
    $p$-values returned by the test for the traditional and \vlne
    energy estimators respectively. A $p$-value above $0.05$ indicates
    that the respective test was passed. 
}
\label{tab:vlne_v3_data_mc_valiadtion}
\end{table}

\subsection{Sensitivity Studies of Searching for a Sterile Neutrino in $\nu_\mu$ Disappearance}

\begin{figure}[h]
  \centering
  \includegraphics[width=0.45\textwidth]{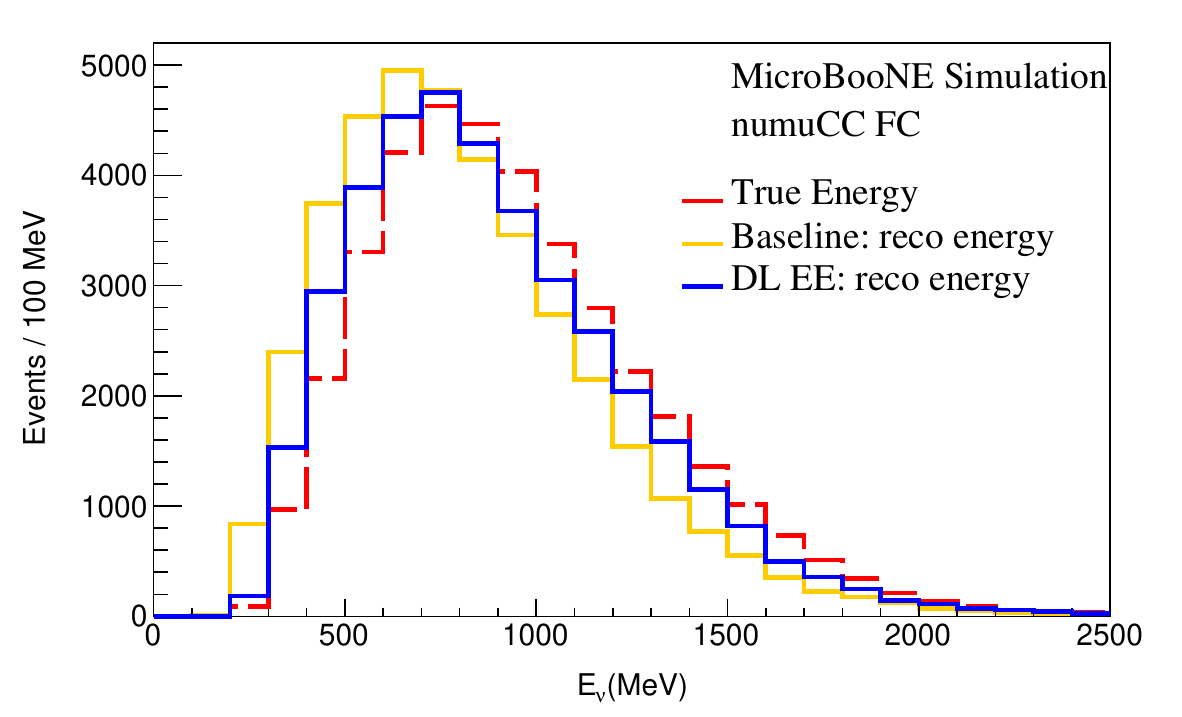}
  \caption{Reconstructed energy spectra of the selected $\nu_\mu$CC FC events assuming no neutrino oscillation 
  in simulation, normalized to $6.369\times10^{20}$ POT.}
  \label{fig:numu_spectrum}
\end{figure}

In this section, we show the impact of the RNN energy estimator on the sensitivity of 
determining neutrino oscillation parameters using the example of a $\nu_\mu$ disappearance search. 
 Assuming $\nu_\mu$ disappearance only, the oscillation probability formula is 
 \begin{equation}\label{eq:numu_sterile}
 P_{\nu_\mu\rightarrow\nu_\mu}=1-\mathrm{sin^2}2\theta_{\mu\mu}\mathrm{sin^2}(\Delta m^2_{41}L/E_{\nu}),
 \end{equation}
 where $L/E_{\nu}$ is the ratio of the neutrino traveling distance and its energy, $\theta_{\mu \mu}$ is the mixing angle that determines the oscillation magnitude, 
 and $\Delta m^2_{41}$ is a mass-squared splitting that modulates the oscillation frequency.
The better energy resolution of the \vlne energy estimator induces less smearing of the oscillating 
features, which leads to a better sensitivity in determining neutrino oscillation parameters.

\begin{figure}[h]
  \centering
  \includegraphics[width=0.45\textwidth]{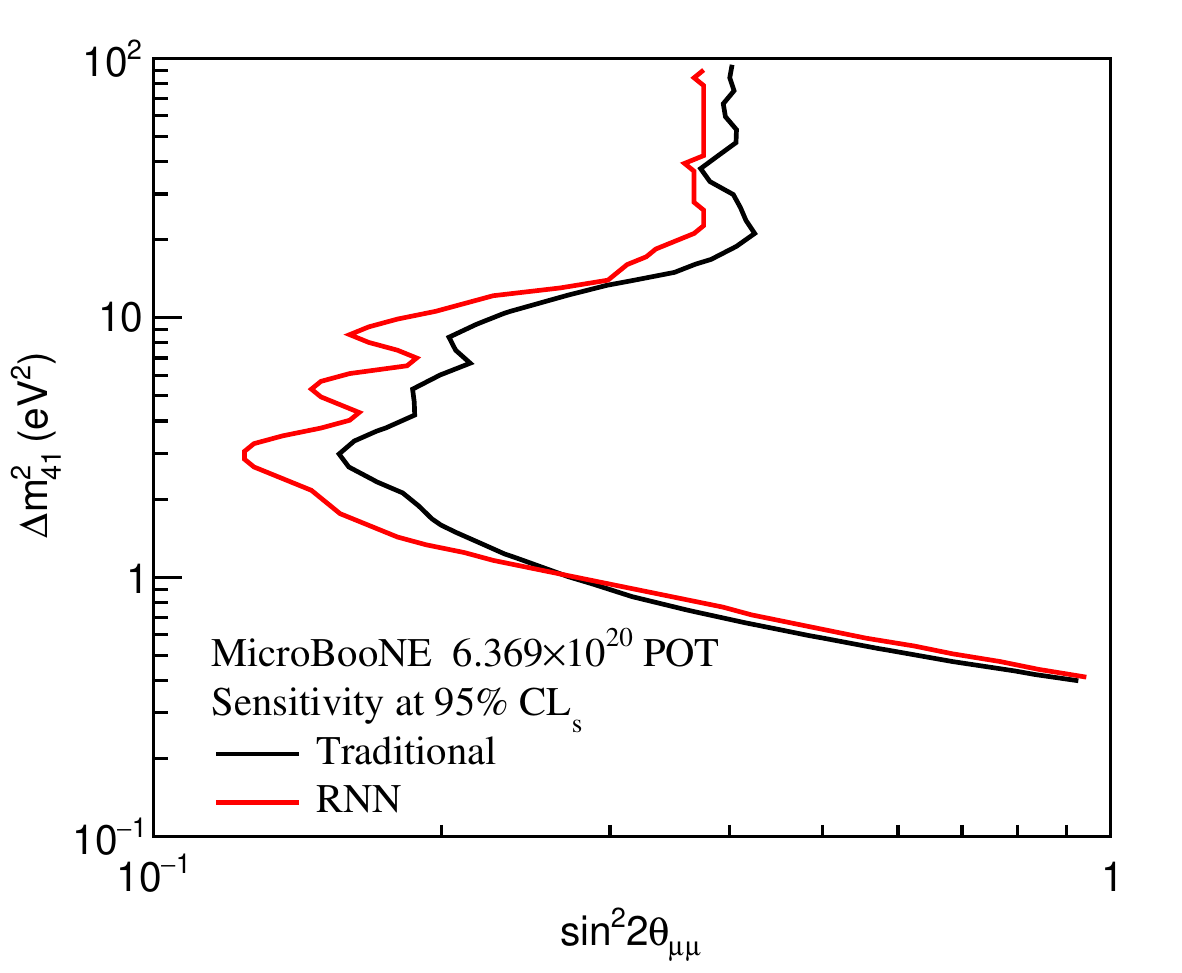}
  \caption{
  MicroBooNE Gaussian $\mathrm{CL_s}$~\cite{MicroBooNE:2022sdp} sensitivity contours at the 95\% CL in the plane of $\mathrm{\Delta m^{2}_{41}}$ and $\mathrm{sin^{2}}\theta_{\mu\mu}$ from the traditional (black) and the RNN (red) methods.
  }
  \label{fig:uu_CLs_sens_95}
\end{figure}

The CC $\nu_{\mu}$-argon interaction event selection and corresponding statistical and 
 systematic uncertainties are taken from Ref.~\cite{MicroBooNE:2021nxr}. 
\Fref{fig:numu_spectrum} shows the reconstructed energy spectra of the the selected $\nu_\mu$ CC 
FC events with no oscillations. Compared to the traditional energy estimator, the \vlne energy estimator
reconstructs a neutrino energy spectrum that is much closer to the truth value.
\Fref{fig:uu_CLs_sens_95} 
shows the sensitivity contours at the 95\% CL in the plane of $\mathrm{\Delta m^{2}_{41}}$ and 
$\mathrm{sin^{2}}2\theta_{\mu\mu}$ for the traditional energy reconstruction and the RNN energy estimator, 
respectively. As expected, the sensitivity is improved especially in the intermediate $\mathrm{\Delta m^{2}_{41}}$ 
region where the oscillation pattern changes considerably over different neutrino energies.
To avoid biases, the study depicted in \fref{fig:uu_CLs_sens_95} does not utilize any training samples from the RNN model.

\section{Conclusions}\label{sec:conclusion}

We developed a deep-learning energy estimation method for charged current neutrino interactions in LArTPC detectors.
This method is based on a recurrent neural network (RNN) architecture.
As inputs it uses reconstructed calorimetry and final-state particle kinematics obtained from the particle flow information.
As outputs, it provides inferences about the energy of the neutrino and the energy of the primary outgoing lepton.
Evaluating this method with simulations, we have shown that the RNN energy estimator's performance is superior to the Wire-Cell traditional energy estimator in terms of bias and resolution.

In order to make this method more resilient to inconsistencies between simulated and real data, we apply the RNN to the reconstructed particle flow information. This is at the cost of inheriting some inefficiencies from the event reconstruction algorithm. In principle, the RNN is capable of correcting for both mis-estimated particle energies and energy from particles that were not reconstructed (either due to inefficiencies or because they do not produce observable signatures).  By training an RNN on only the Wire-Cell traditional neutrino energy estimate, we can determine to what extent the RNN leverages individual particle information to determine the missing energy contribution.  In this case, the performance is better than the traditional calculation, but falls short compared to the full RNN, indicating that the RNN can indeed learn correlations between the reconstructed particles and the missing energy.

Besides performance evaluations with simulations, 
a set of dedicated model validation tests was performed to demonstrate that the \OverallModel enhanced by the RNN energy 
estimator is compatible with MicroBooNE experimental data within the model uncertainties. 
Using a simple example of searching for a sterile neutrino with $\nu_\mu$ disappearance oscillations, we show the impact of this RNN energy estimator on the physics sensitivity. Besides the impact on neutrino oscillations, we expect that this technique can be extended to cross-section measurements in both inclusive and exclusive interaction channels. Adaptation of this method for other LArTPC experiments, such as DUNE~\cite{DUNE:2021tad} and SBND~\cite{Machado:2019oxb}, is underway.

\begin{acknowledgments}
This document was prepared by the MicroBooNE collaboration using the
resources of the Fermi National Accelerator Laboratory (Fermilab), a
U.S. Department of Energy, Office of Science, HEP User Facility.
Fermilab is managed by Fermi Research Alliance, LLC (FRA), acting
under Contract No. DE-AC02-07CH11359.  MicroBooNE is supported by the
following: 
the U.S. Department of Energy, Office of Science, Offices of High Energy Physics and Nuclear Physics; 
the U.S. National Science Foundation; 
the Swiss National Science Foundation; 
the Science and Technology Facilities Council (STFC), part of the United Kingdom Research and Innovation; 
the Royal Society (United Kingdom); 
the UK Research and Innovation (UKRI) Future Leaders Fellowship; 
and the NSF AI Institute for Artificial Intelligence and Fundamental Interactions. 
Additional support for 
the laser calibration system and cosmic ray tagger was provided by the 
Albert Einstein Center for Fundamental Physics, Bern, Switzerland. We 
also acknowledge the contributions of technical and scientific staff 
to the design, construction, and operation of the MicroBooNE detector 
as well as the contributions of past collaborators to the development 
of MicroBooNE analyses, without whom this work would not have been 
possible. For the purpose of open access, the authors have applied 
a Creative Commons Attribution (CC BY) public copyright license to 
any Author Accepted Manuscript version arising from this submission.

The initial development of the \vlne energy estimator occurred at the NOvA experiment. We are grateful to the NOvA collaboration for multiple discussions and suggestions to improve the \vlne energy estimator. 
\end{acknowledgments}

\appendix
\section{TRAINING DETAILS}
\label{sec:app:train_details}

The \vlne energy estimator is trained with the help of the \texttt{vlne}~\cite{vlne_software}
package, which is built on top of the TensorFlow/Keras (v2.9) frameworks.
The training is performed for up to 200 epochs using the \texttt{RMSprop} optimizer, and 
an initial learning rate of $0.001$. The learning rate is progressively annealed by the
\texttt{ReduceLROnPlateau} scheduler (\texttt{patience = 5, factor = 0.5}). 

The \vlne training is terminated if the validation loss has not improved over the last
40 epochs, or if the total number of training epochs has reached 200. To reduce
overfitting, an $L_2$ regularization is applied with a strength of $0.008$.
As an additional regularization technique we have randomly shuffled the order
of particles in each event.

For the training we have used a batch size of 1024. When updating the neutrino energy
spectrum with flat event weights we applied the weights only to the the neutrino part
$\mathcal{L}_\nu$ of the loss function.
We did not reweight the primary lepton part $\mathcal{L}_{\text{lep}}$ of the loss.
The training is run on a single NVIDIA GeForce RTX 3090 GPU. Because of the reliance 
on high-level input variables, it takes less than 20 minutes to train a single \vlne energy estimator.

To determine the final network configuration we have performed several hyperparameter
sweeps. We have explored a grid of the following parameters:
learning rate, number of training epochs, number of features in the LSTM cell and fully connected layers,
depth of the fully connected layers, regularization type ($L_1$ vs $L_2$), and the strength of the regularization.
\section{\vlne FOR $\nu_e$CC ENERGY ESTIMATION}
\label{sec:app_nue}

In this appendix we explore the basic performance of the \vlne energy
estimator on the $\nu_e$ events.
The training sample of the \vlne energy estimator contains charged-current 
electron neutrino interaction ($\nu_e$ CC) events.
Therefore, the \vlne is capable of predicting energies of such events as well. Because of the
small statistics of the $\nu_e$ CC sample we are unable to perform a similar data/MC
validation as that with the $\nu_\mu$ CC sample. This situation is expected 
to be improved with future experiments.

\begin{figure}[tbh]
  \centering
  \includegraphics[width=0.45\textwidth]{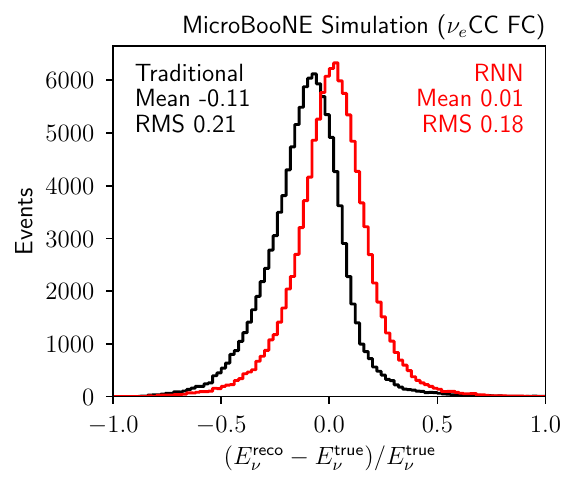}
  \caption{
    $\nu_e$ energy resolution histograms for the traditional \uboone
    energy estimator (black) and the reweighted training of the \vlne energy
    estimator (red) in the Fully Contained (FC) CC sample.
  }
  \label{fig:vlne_new_nue_fc_resolution_total}
\end{figure}

\begin{figure}[tbh]
  \centering
  \includegraphics[width=0.45\textwidth]{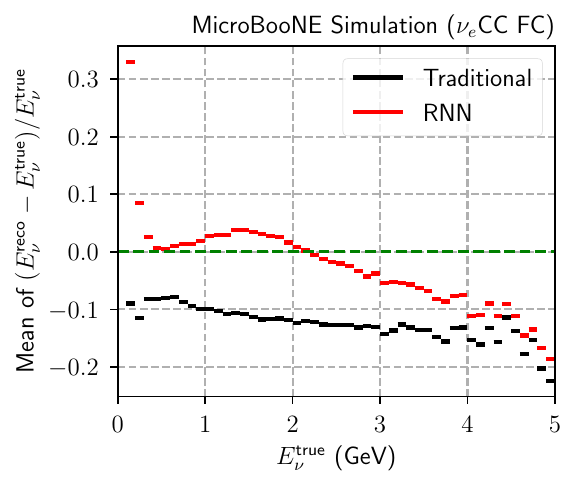}
  \caption{
    $\nu_e$ energy resolution bias for the traditional \uboone
    energy estimator (black) and the reweighted training of the \vlne energy
    estimator (red) in the Fully Contained (FC) CC sample.
  }
  \label{fig:vlne_new_nue_fc_binstat_total_mean}
\end{figure}

\Fref{fig:vlne_new_nue_fc_resolution_total} shows the $\nu_e$ CC energy resolution
histogram of the \vlne compared to the traditional energy estimator. In terms of
the RMS, the \vlne slightly outperforms the traditional energy estimator
($18 \%$ vs $21 \%$).
\Fref{fig:vlne_new_nue_fc_binstat_total_mean} illustrates biases of the \vlne and
traditional energy estimators. For the majority of true energy bins, the \vlne displays
a much smaller bias compared to the traditional energy estimator. Only at the lowest
energy bin does the \vlne acquires a large bias.

\begin{figure}[tbh]
  \centering
  \includegraphics[width=0.45\textwidth]{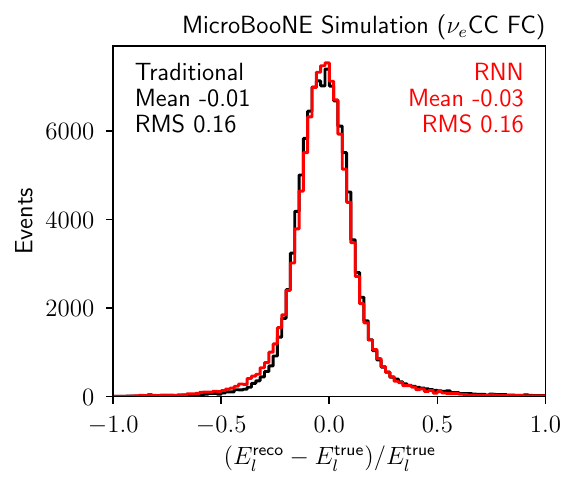}
  \caption{
    Primary electron energy resolution histograms for the traditional \uboone
    energy estimator (black) and the reweighted training of the \vlne energy
    estimator (red) in the Fully Contained (FC) CC sample.
  }
  \label{fig:vlne_new_nue_fc_resolution_primary}
\end{figure}

\begin{figure}[tbh]
  \centering
  \includegraphics[width=0.45\textwidth]{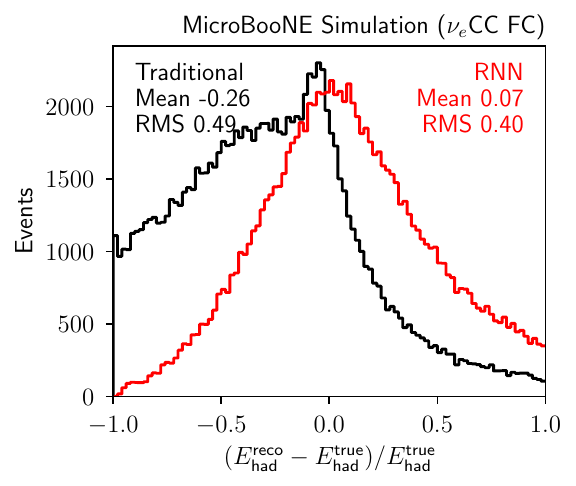}
  \caption{
    $\nu_e$CC hadronic energy resolution histograms for the traditional \uboone
    energy estimator (black) and the reweighted training of the \vlne energy
    estimator (red) in the Fully Contained (FC) CC sample.
  }
  \label{fig:vlne_new_nue_fc_resolution_secondary}
\end{figure}

\Fref{fig:vlne_new_nue_fc_resolution_primary} and \fref{fig:vlne_new_nue_fc_resolution_secondary}
show the energy resolution histograms of the primary lepton and hadronic energies, respectively.
There is little difference in the resolution of the energy of the primary lepton.
However, the \vlne achieves superior hadronic energy reconstruction, with improvement in RMS
from $49 \%$ to $40 \%$.

\section{ENERGY ESTIMATION OF THE PARTIALLY CONTAINED EVENTS}
\label{sec:app_pc_events}

\begin{figure*}[tbh]
  \centering
  \captionsetup[subfigure]{justification=centering}
  \begin{subfigure}[t]{0.45\textwidth}
    \centering
    \includegraphics[width=\textwidth]{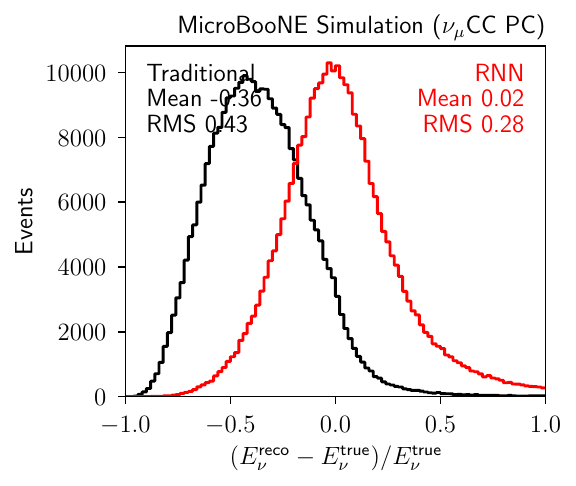}
    \caption{$\nu_\mu$ CC Energy Resolution}
  \end{subfigure}
    \begin{subfigure}[t]{0.45\textwidth}
    \includegraphics[width=\textwidth]{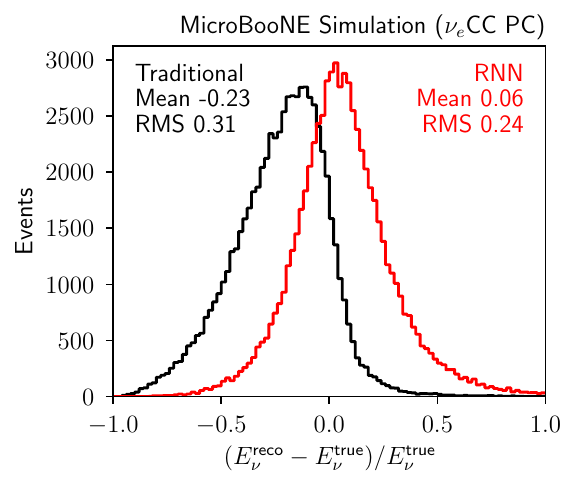}
    \caption{$\nu_e$ CC Energy Resolution}
  \end{subfigure}
  \caption{
    $\nu_\mu$ and $\nu_e$ energy resolution histograms for the traditional
    \uboone energy estimator (black) and the reweighted training of the
    \vlne energy estimator (red), evaluated on the samples of PC CC events.
  }
  \label{fig:vlne_new_pc_resolution_total}
\end{figure*}

In this appendix, we review the performance of the \vlne energy estimator on the
Partially Contained (PC) events and compare it to the performance of the traditional
\uboone energy estimator.
\Fref{fig:vlne_new_pc_resolution_total} compares neutrino energy 
resolution for the $\nu_\mu$ CC and $\nu_e$ CC events.
For the PC events, the \vlne energy estimator demonstrates improvement in the energy resolution from $43\%$ to $28\%$ for $\nu_\mu$ CC, and improvement from $31\%$ to $24\%$ for $\nu_e$ CC, compared to the traditional \uboone energy estimator.

\begin{figure*}[tbh]
  \centering
  \captionsetup[subfigure]{justification=centering}
  \begin{subfigure}[t]{0.45\textwidth}
    \centering
    \includegraphics[width=\textwidth]{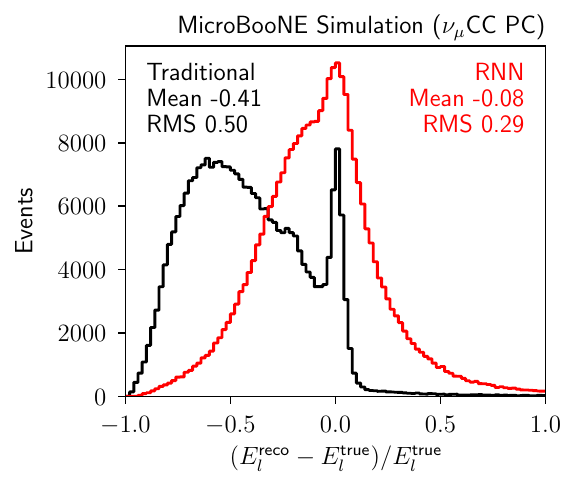}
    \caption{Primary $\mu$ Energy Resolution}
  \end{subfigure}
    \begin{subfigure}[t]{0.45\textwidth}
    \includegraphics[width=\textwidth]{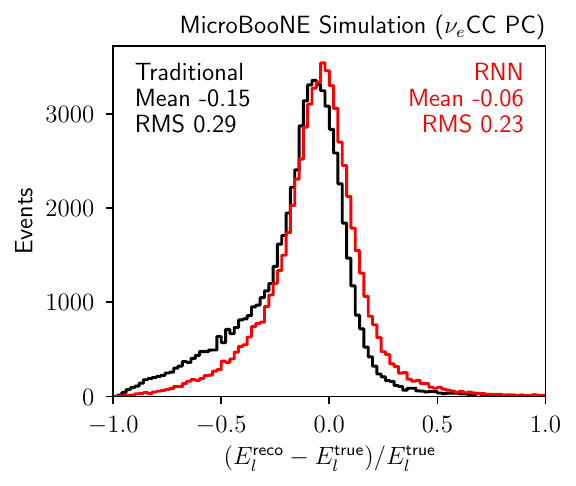}
    \caption{Primary $e$ Energy Resolution}
  \end{subfigure}
  \caption{
    Primary muon and electron energy resolution histograms for the traditional
    \uboone energy estimator (black) and the reweighted training of the
    \vlne energy estimator (red), evaluated on the samples of PC CC events.
  }
  \label{fig:vlne_new_pc_resolution_primary}
\end{figure*}

Likewise, \fref{fig:vlne_new_pc_resolution_primary} compares energy resolutions of
the primary leptons in the $\nu_\mu$ and $\nu_e$ CC events. It shows a large improvement
in reconstructing the energy of the primary muons with the RMS improving from $50\%$ for
the traditional \uboone energy estimator down to $29\%$ for the \vlne energy estimator.

The improvement for the energy reconstruction of the primary electron energy is much smaller, 
with the RMS improving from $29\%$ down to $23\%$. Moreover, the majority
of this improvement comes from the tail of the distribution, indicating that the
traditional electron energy reconstruction approach is close to the optimal.

\begin{figure*}[tbh]
  \centering
  \captionsetup[subfigure]{justification=centering}
  \begin{subfigure}[t]{0.45\textwidth}
    \centering
    \includegraphics[width=\textwidth]{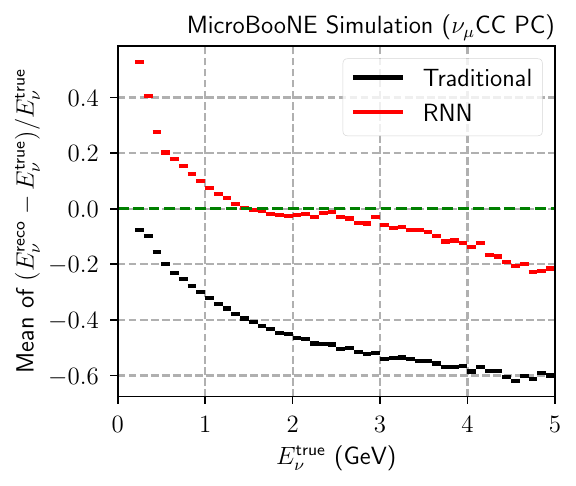}
    \caption{$\nu_\mu$ Energy Resolution Bias}
  \end{subfigure}
    \begin{subfigure}[t]{0.45\textwidth}
    \includegraphics[width=\textwidth]{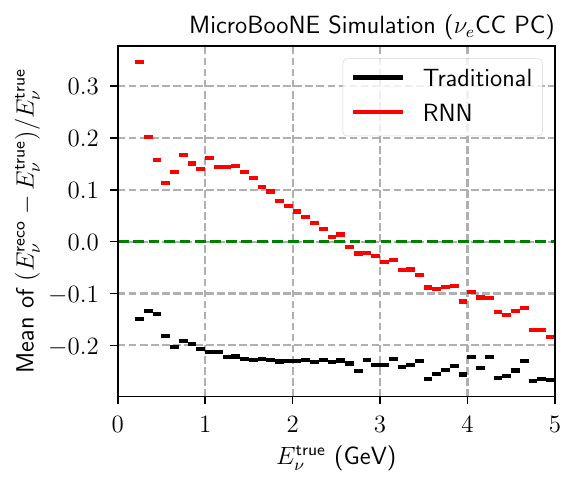}
    \caption{$\nu_e$ Energy Resolution Bias}
  \end{subfigure}
  \caption{
    $\nu_\mu$ and $\nu_e$ energy resolution bias plot for the traditional
    \uboone energy estimator (black) and the reweighted training of the
    \vlne energy estimator (red), evaluated on the samples of PC CC events.
  }
  \label{fig:vlne_new_pc_bias_total}
\end{figure*}

Finally, \fref{fig:vlne_new_pc_bias_total} shows the bias of the neutrino
energy estimators as a function of the true neutrino energy. The $\nu_\mu$ CC energy
bias of the \vlne energy estimator is smaller than the bias of the traditional \uboone
energy estimator for the majority of neutrino energies.
It is, however, larger at small true $E_\nu$ ($E_\nu < \SI{0.5}{GeV}$).
This behavior is similar to the energy bias
behavior for the FC events, observed in \sref{sec:train_initial}. The energy
reweighting reduces the low energy bias for the FC events, but evidently, it is not 
sufficient to remove the bias for the PC events. For the $\nu_e$ energy estimation, 
\fref{fig:vlne_new_pc_bias_total} shows a different
behavior. The \vlne energy estimator has a smaller bias for virtually the entire energy
range, compared to the traditional \uboone energy estimator. However, it also exhibits
large oscillations in the low energy region.

\section{RESOLUTION DEPENDENCE ON ENERGY}
\label{sec:app_resolution_vs_energy}

\begin{figure*}[tbh]
  \centering
  \captionsetup[subfigure]{justification=centering}
  \begin{subfigure}[t]{0.45\textwidth}
    \centering
    \includegraphics[width=\textwidth]{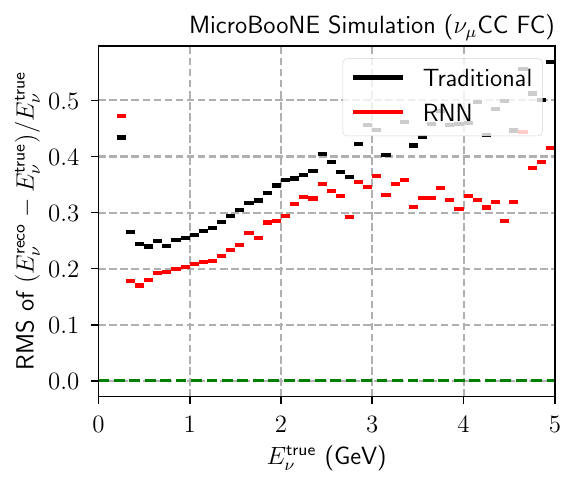}
    \caption{$\nu_\mu$ FC}
  \end{subfigure}
    \begin{subfigure}[t]{0.45\textwidth}
    \includegraphics[width=\textwidth]{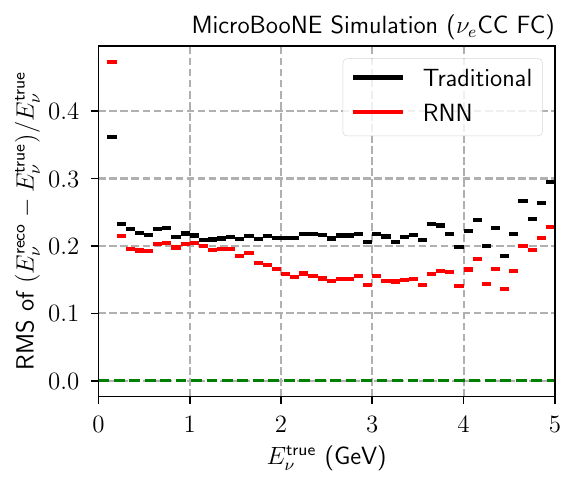}
    \caption{$\nu_e$ FC}
  \end{subfigure}  
  \begin{subfigure}[t]{0.45\textwidth}
    \centering
    \includegraphics[width=\textwidth]{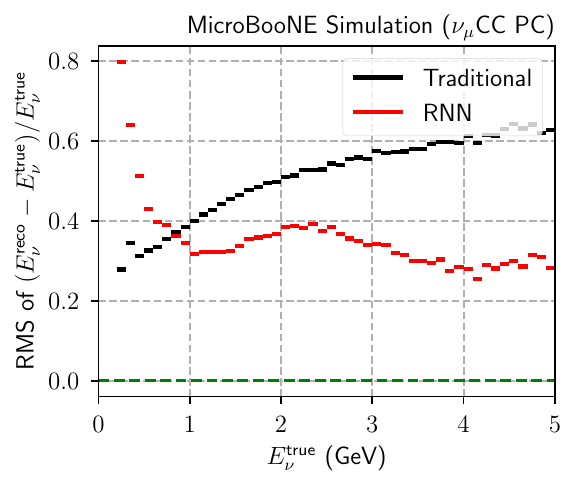}
    \caption{$\nu_\mu$ PC}
  \end{subfigure}
    \begin{subfigure}[t]{0.45\textwidth}
    \includegraphics[width=\textwidth]{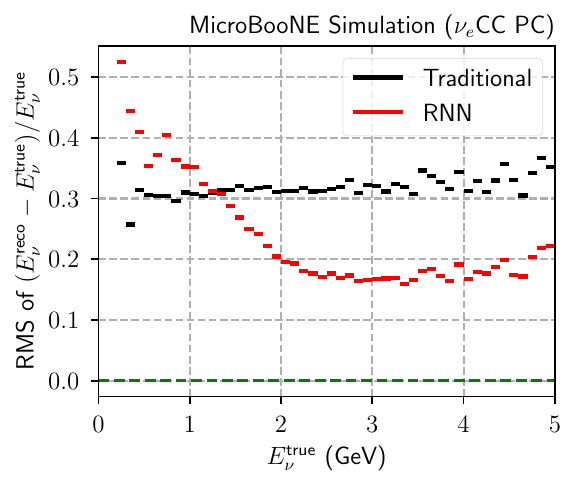}
    \caption{$\nu_e$ PC}
  \end{subfigure}
  \caption{
    $\nu_\mu$ and $\nu_e$ energy resolution vs true neutrino energy plots for
    the traditional \uboone energy estimator (black) and the reweighted
    training of the \vlne energy estimator (red), evaluated on the samples
    of FC and PC events.
  }
  \label{fig:vlne_new_rms_vs_true_total}
\end{figure*}

In this appendix, we review the binned statistics plots, depicting energy resolution
versus true neutrino energy bins. \Fref{fig:vlne_new_rms_vs_true_total} illustrates
the behavior of the RMS of the energy resolution as a function of the true neutrino
energy. For the FC events, the \vlne energy estimator has better neutrino energy resolution compared to the traditional \uboone energy estimator across the entire energy range.
For the PC events, however, the \vlne loses its performance in the low-energy region.

\bibliographystyle{apsrev4-2}
\bibliography{main}

\end{document}